\documentclass[final,5p,times,twocolumn,authoryear]{elsarticle}

\usepackage{amssymb}
\usepackage{array}
\usepackage{amsmath,amsfonts}
\usepackage{subfigure}

\usepackage{textcomp}
\usepackage{stfloats}
\usepackage{url}
\usepackage{verbatim}
\usepackage{booktabs} 
\usepackage{tabularx} 
\usepackage{color,soul} 
\usepackage[normalem]{ulem} 

\usepackage{graphicx}
\usepackage{balance}
\usepackage[linesnumbered,ruled,vlined]{algorithm2e}
\usepackage{float}
\usepackage[utf8]{inputenc}

\usepackage{lineno}

\newif\ifhighlightrevision
\highlightrevisionfalse 

\newenvironment{rev}{
  \color{black}
}{
  \color{black}
}

\journal{Machine Learning with Applications (VSI: Autonomous Systems)}

\begin{document}

\begin{frontmatter}



\title{Multi-source Plume Tracing via Multi-Agent Reinforcement Learning}


\author[inst1]{Pedro Antonio Alarcon Granadeno \corref{cor1}}
\ead{palarcon@nd.edu}
\cortext[cor1]{Corresponding Author}

\author[inst1]{Theodore Chambers}
\ead{tchambe2@nd.edu}

\author[inst1]{Jane Cleland-Huang}
\ead{janeClelandHuang@nd.edu}

\affiliation[inst1]{organization={University of Notre Dame, Computer Science And Engineering},
            addressline={384 Fitzpatrick Hall of Engineering}, 
            city={Notre Dame},
            postcode={46556}, 
            state={IN},
            country={USA}}

\begin{abstract}
\begin{rev}
Industrial catastrophes like the Bhopal disaster (1984) and the Aliso Canyon gas leak (2015) demonstrate the urgent need for rapid and reliable plume tracing algorithms to protect public health and the environment. Traditional methods, such as gradient-based or biologically inspired approaches, often fail in realistic, turbulent conditions. To address these challenges, we present a Multi-Agent Reinforcement Learning (MARL) algorithm designed for localizing multiple airborne pollution sources using a swarm of small uncrewed aerial systems (sUAS). Our method models the problem as a Partially Observable Markov Game (POMG), employing a Long Short-Term Memory (LSTM)-based Action-specific Double Deep Recurrent Q-Network (ADDRQN) that uses full sequences of historical action-observation pairs, effectively approximating latent states. Unlike prior work, we use a general-purpose simulation environment based on the Gaussian Plume Model (GPM), incorporating realistic elements such as a three-dimensional environment, sensor noise, multiple interacting agents, and multiple plume sources. The incorporation of action histories as part of the inputs further enhances the adaptability of our model in complex, partially observable environments. Extensive simulations show that our algorithm significantly outperforms conventional approaches. Specifically, our model allows agents to explore only 1.29\% of the environment to successfully locate pollution sources. 
\end{rev}
\end{abstract}




\end{frontmatter}


\section{Introduction}
\begin{rev}
The Bhopal disaster in 1984, one of the most severe industrial 
catastrophes in history, resulted from the release of methyl 
isocyanate gas, leading to thousands of immediate fatalities and long-term health repercussions for hundreds of thousands more
\citep{Broughton2005}. Similarly, the Aliso Canyon gas leak in 2015, the 
largest methane release in U.S. history, emitted over 100,000 tons of 
methane over several months, causing substantial environmental harm, 
contributing significantly to greenhouse gas emissions, and 
necessitating the evacuation of thousands of residents due to health 
hazards \citep{doi:10.1126/science.aaf2348}. These incidents underscore the urgent need for 
prompt pollutant detection during an emergency response to manage and 
mitigate toxic airborne plumes for  environmental and public health 
protection. 

Plume tracing -- the process of identifying the point source(s) of a chemical discharge dispersed through some fluid medium, such as air or water -- is a key task across multiple domains ranging from environmental conservation to national security. Notable real-world applications include the localization of oil spills \citep{1470515}, detection of gas leaks \citep{GasSourceMicro}, determination of fire 
origins \citep{1470520}, identification of explosive ordnance 
\citep{10.1117/12.719742}, and monitoring of environmental pollution \citep{BAYAT201776}. In recent years, small Uncrewed Aerial Systems (sUAS) have emerged as valuable tools for plume tracing. UAVs offer significant advantages through remote sensing and the ability to explore hazardous environments. Their mobility allows for more effective localization of chemical plume sources compared to static 
systems. Multiple sUAS can operate in coordination, covering larger areas more quickly, sharing data in real time, and adaptively responding to dynamic plume conditions.

Traditional approaches to plume tracing often model the problem as a nonlinear optimization task within a chemical field, relying on gradient-based methods that follow trajectories toward high chemical concentrations. While these methods perform adequately in stable conditions with continuous concentration gradients, they falter in realistic scenarios characterized by turbulent eddy motions and advection mechanisms. Such complexities result in high variability of plume shapes, intermittent chemical concentrations, and discontinuities, rendering gradient methods unreliable. Biologically inspired strategies, though efficient in nature, often lack the flexibility to generalize across various plume forms and conditions. These strategies typically require extensive oversampling of the search area, leading to redundant movements and inefficient resource utilization, especially in emergency scenarios where rapid and efficient source localization is crucial. Recent advancements in reinforcement learning (RL) offer promising 
alternatives. However, existing RL solutions  are typically designed for single-source localization, employ single-agent frameworks, or operate under simplified assumptions about plume characteristics, limiting their applicability in complex, real-world scenarios involving diverse plume shapes, multiple local maxima, and three-dimensional maneuverability.

To address the limitations of traditional plume tracing methods and to leverage UAV advantages, this paper proposes a Multi-Agent Reinforcement Learning Algorithm (MARL) for localizing multiple airborne pollution plumes. Using the Gaussian Dispersion Model, we construct realistic 3-D pollution environments and model the problem as a Partially Observable Markovian/Stochastic Game (POMG) within a multi-agent cooperative framework. Our approach introduces Action-specific Double Deep Recurrent Q-Networks (ADDRQN), which incorporate action histories to enhance decision-making and localization accuracy. By integrating Long Short-Term Memory (LSTM) \citep{Graves2012} networks, we effectively manage partial observability challenges. We substantiate the efficacy of our approach through comparative analysis with several baseline techniques, including DDRQN, DRQN, and DQN. To the best of our knowledge, this research is the first work to propose a multi-agent reinforcement learning solution for 3D plume tracing of multiple airborne pollutants. 
\end{rev}
Our main contributions are as follows:

\begin{enumerate}
    \item We articulate our plume tracing problem as a Partially Observable Markov Game (POMG) in the context of Multi-Agent Reinforcement Learning (MARL).
    \item We design an LSTM-based Action-Specific Double Deep Recurrent Network (ADDRQN) that explicitly incorporates actions within historical observations, resulting in a learning mechanism that enhances adaptability in partially observable scenarios. 
    \item We design a diverse and complex realistic plume dispersion training environment that features three-dimensional space, multiple plume sources, and intermittent pollution patterns tailored for multi-agent training scenarios. Research in Gaussian Models indicate this simulation archetype has broad applicability. 
\end{enumerate}

The remainder of the paper is structured as follows. In section \ref{sec:related-work}, we discuss the limitations of current related state-of-the-art plume tracing methods and highlight the potential for reinforcement learning as a viable surrogate. Section \ref{sec:prob_form} outlines the problem space of plume tracing tasks and details the key roles and capabilities that UAVs assume within this context. Section \ref{sec:GPM} describes the Gaussian Plume Dispersion model used to simulate and generate three-dimensional, multi-point source plumes. Section \ref{sec:modeling} introduces the concepts of Partially Observable Markov Games (POMGs) and Recurrent Reinforcement Learning (RRL). Section \ref{sec:ADDRQN}  and \ref{sec:Exp_design} introduces our ADDRQN MARL framework and details its training methodology. Section \ref{sec:Performance} designs an experimental test-bed and presents the performance of ADDRQN against other established algorithms. Section \ref{sec:limitations} discussed limitations of our proposed framework and suggest future improvements. Finally, Section \ref{sec:conclusions} concludes our work.

\section{Related Work}\label{sec:related-work}

\subsection{Gradient-Based Plume Tracing Methods}
\begin{rev}
Traditional plume tracing approaches often model the problem as a nonlinear optimization task within a chemical field, using gradient-based methods to follow trajectories toward areas of high chemical concentration. Neumann et al. \citep{GasSourceMicro} developed a pseudo-gradient method by combining surge anemotaxis and chemotaxis. In their work, a micro-drone gathers wind and gas sensor data, then moves orthogonally to the wind direction, calculating its next position using gradient information from these measurements. Mayhew et al. \citep{12} implemented a hybrid controller based on consecutive minimizations along a line directed by conjugate vectors. Yungaicela-Naula et al. \citep{8453430} presented a gradient-based algorithm with a Bayesian probabilistic method for pollutant source discovery.

These methods are appealing due to their simplicity and effectiveness in stable, continuous concentration gradients. However, real-world chemical propagation is often irregular because of turbulent eddy motions and advection mechanisms, leading to plumes with intermittent concentrations and discontinuities. While gradient-based methods are computationally efficient and straightforward to implement, their reliance on continuous and smooth gradients limits their applicability in turbulent environments. This limitation motivates the search for alternative strategies capable of handling the stochastic nature of plume dispersion.
\subsection{Biologically-Inspired Plume Tracing Strategies}

Another line of research draws inspiration from natural search mechanisms. Li et al. \citep{1618529} proposed a zigzag-pattern search strategy inspired by moths moving upwind along pheromone plumes. Hayes et al. \citep{1021067} used an outward spiral pattern for local plume detection, followed by a surge algorithm for plume traversal. Alvear et al. \citep{Alvear2018} combined chemotaxis meta-heuristics with adaptive spiral mobility patterns to identify areas of maximum pollution. Farrell et al. \citep{1522521} took inspiration from olfactory-based foraging behaviors observed in lobsters and crabs to develop a plume tracing algorithm for small unmanned aerial systems.

While these biologically-inspired methods introduce robustness by mimicking evolved natural behaviors, they often lack adaptability to environments significantly different from those in which the biological organisms operate. These behavior-based strategies tend to be too rigid to generalize to various plume forms, making their performance heavily dependent on specific conditions. Predetermined patterns like spirals or zigzags may not be optimal in complex urban landscapes or cluttered environments where obstacles and changing wind conditions can disrupt search patterns. Additionally, these strategies often require oversampling the search area, leading to redundant movements and inefficient resource utilization. In emergency scenarios where rapid and efficient source localization is crucial, these limitations become particularly problematic. This underscores the need for more flexible and adaptive strategies capable of navigating diverse and unpredictable environments efficiently.

\subsection{Reinforcement Learning in Plume Tracing}

Given the limitations of gradient-based and behavior-driven methods, reinforcement learning (RL) has emerged as a viable alternative. The success of RL in challenges like Atari games \citep{mnih2013playing} and AlphaGo \citep{Silver2016, Silver2017} has spurred its application across various sectors, including autonomous driving and algorithmic trading.

In environmental monitoring, plume tracing tasks have begun to reap the benefits of contemporary RL research. Tate et al. \citep{9706111} employed a Q-learning algorithm enabling an agent to discern multi-parameter gradient relationships between methane and oxygen for underwater chemical source localization. However, their study focuses on a single-agent framework without noise tolerance and is limited to one training environment, which restricts its generalizability.

Similarly, Hu et al. \citep{8598800} designed an LSTM-based Deep Deterministic Policy Gradient (DDPG) algorithm to train an agent in partially observable, turbulent underwater environments modeled via Reynolds-averaged Navier-Stokes equations. While their approach addresses turbulence, it operates in a two-dimensional space and does not consider multi-agent collaboration, limiting its applicability in more complex scenarios.

Mohammed et al. \citep{s22166118} introduced an LSTM-based Deep Q-Network (DQN) trained on a 2D pollution environment simulated through Gaussian distribution and kriging interpolations. Although their model efficiently localized multiple zones exceeding a predefined Air Quality Index threshold—achieving 28\% faster performance than a spiral-behavior strategy—it is confined to a two-dimensional space, does not incorporate sensor noise, and lacks a multi-agent framework.

\subsection{Challenges in Existing RL Approaches and Our Proposed Solution}

The aforementioned RL-based studies have contributed valuable insights but exhibit limitations that hinder their applicability to more general plume tracing tasks. The absence of multi-agent frameworks prevents the exploration of cooperative strategies that could enhance search efficiency. Operating primarily in two-dimensional spaces overlooks the complexity introduced by three-dimensional environments, which is especially relevant for aerial drones or underwater vehicles operating at varying depths. Additionally, the lack of sensor noise modeling reduces the robustness of these approaches when deployed in real-world conditions where measurements are often noisy and uncertain.

Our approach addresses these gaps by leveraging a multi-agent reinforcement learning framework that integrates joint actions into the learning process. Operating in a complex environment featuring three dimensions, intermittent plume concentrations, state noise, and multiple sources, our system offers a more realistic and scalable solution for plume tracing tasks. By enabling multiple agents to coordinate their actions, we achieve more efficient exploration of the search space and faster localization of sources. Incorporating three-dimensional maneuverability and accounting for environmental noise enhance the applicability of our approach to real-world scenarios.

We employ the Gaussian Dispersion Model (GDM), a widely recognized tool for simulating the dispersion of various pollutants -- including sulfur dioxide, nitrogen oxides, particulate matter, carbon monoxide, volatile organic compounds, ammonia, and hydrogen sulfide. Known for its simplicity and computational efficiency, the GDM effectively models short to medium-range dispersion over flat, homogeneous terrains with continuous emissions under neutral and stable atmospheric conditions. This allows our method to be adaptable across different environmental settings, not limited to specific applications like underwater methane detection or air quality monitoring without source tracking. Moreover, we conduct robust testing using unseen environments to provide quantitative performance metrics. Our method operates over a large state space of $16 \times 16 \times 16 \times  20 = 81,920$ states, significantly exceeding the state spaces considered in previous RL applications to plume tracing.  By addressing the limitations of existing RL approaches and offering a robust framework adaptable across different environmental settings, our method lays the groundwork for future research and practical implementations in environmental monitoring and emergency response situations
\end{rev}
\section{Problem Formulation} 
\label{sec:prob_form}

Plume tracing strategies are typically decomposed into three distinct subroutines, namely, plume detection, plume traversal and source confirmation. In the initial stage, plume detection is carried out through a detection-threshold mechanism, wherein a concentration exceeding a predetermined level ascertains the presence of a plume within the monitored area. Successful plume detection initiates a search strategy, in which sUAS strategically follow a direction that leads to higher concentrations of a substance towards the source, effectively narrowing down the search area. Finally, upon approaching a potential source, the tracing strategy concludes with a verification process either via visual imagery, human-in-the-loop, or other feasible means. Our Multi-Agent Reinforcement Learning methodology is capable of autonomously executing plume detection and tracing to source as integrated components of its trained policy; however, source verification algorithms is outside the scope of this study.

In a typical scenario, a swarm of sUAS is strategically deployed near the vicinity of several pollution source emittents. These sUAS possess the capability to navigate through three-dimensional space via a set of predefined actions \( \mathcal{A} \). We presume that each $i$th sUAS is equipped with an array of on-board sensors that capture instantaneous geolocation $p_{t}^{i} = (x_{t}^{i},y_{t}^{i},z_{t}^{i})$ and its respective pollution concentration $C(p_{t}^{i})$ at time t,  relative to a designated coordinate frame of reference.  The sensors, subject to varying degrees of precision, can influence the accuracy of these concentration readings due to inherent imperfections. The observed, potentially noisy, environmental conditions at time \( t \) for the \( i^{th} \) sUAS are denoted as \( o_t^{i} = (C(p_{t}^{i}), p_{t}^{i}) \).  Furthermore, we assume that the sUAS contain mesh radio communication modules -- or other reasonable alternatives -- that enable reliable and robust peer-to-peer communication protocols. Consequently, each sUAS maintains a real-time cumulative record \( H_{t}^{i} \) that encapsulates the sequential environmental observation history and its actions up to $t-1$, expressed as \( H_{t}^{i} = \{ (o_{1}^{i}, a_{0}^{i}), \ldots, (o_{t}^{i}, a_{t-1}^{i}) \} \). During each time step \( t \), all sUAS exchange their individual logs, resulting in a collective history log that aggregates $N$ sUAS records, \( H_{t} = \{ H_{t}^{1}, H_{t}^{2}, \ldots, H_{t}^{N} \} \). This framework defines a search strategy as a navigational policy, $\pi$, mapping collective historical data to specific actions for each sUAS, encapsulated by the function, $\pi : H_t \mapsto a^i_t$, where $a^i_t \in \mathcal{A}$. This mathematical framework serves as the foundational structure for training our plume tracing policies. Further details are explored in subsequent sections.

\section{Gaussian Plume Dispersion Model}
\label{sec:GPM}

Atmospheric dispersion refers to the diffusion of nonreactive pollutants under the influence of turbulent eddy motion and advection caused by meteorological conditions. In particular, Gaussian models (i.e. AERMOD, OCD, GPM, CTDMPLUS, etc.) have been widely adopted by EPA and other industrial operations as a standard approach for simulating the short-range transport of airborne pollutants. Gaussian Plume-based solutions have been extensively studied in a wide range of applications including multi-agent gas source detection \citep{WangGasMulti2d}, modeling realistic pollution in dense roadway interchanges \citep{MIDDLETON19791039}, and identifying livestock facility locations \citep{LIVESTOCKPLUME}. 

The formula that governs atmospheric dispersion is described by the time dependent advection-diffusion equation:

 \begin{equation} \label{eq:advection-difussion equation}
    \frac{\partial C}{ \partial t} =  \nabla K \cdot \nabla C - \nabla \cdot u C + S
\end{equation}

\begin{figure}[!htb]
    \centering
    \subfigure[Three-dimensional Scatter Plot of Plume.]{
        \includegraphics[width=.45\columnwidth]{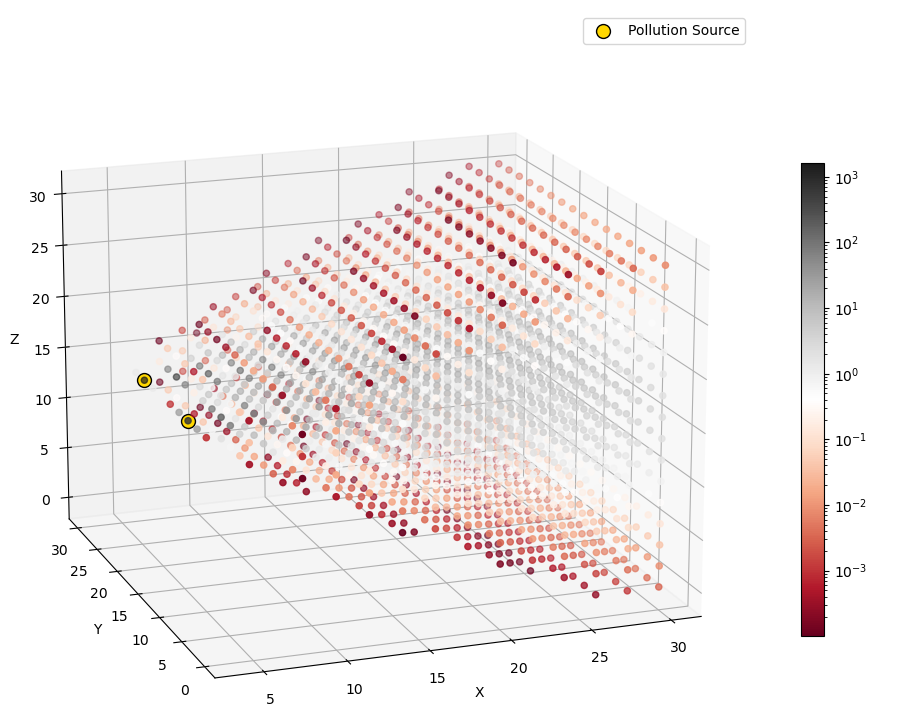}
        \label{fig:scatter-plot}
    }
    \subfigure[Contour plot at specific height (z=14 meters).]{
        \includegraphics[width=.45\columnwidth]{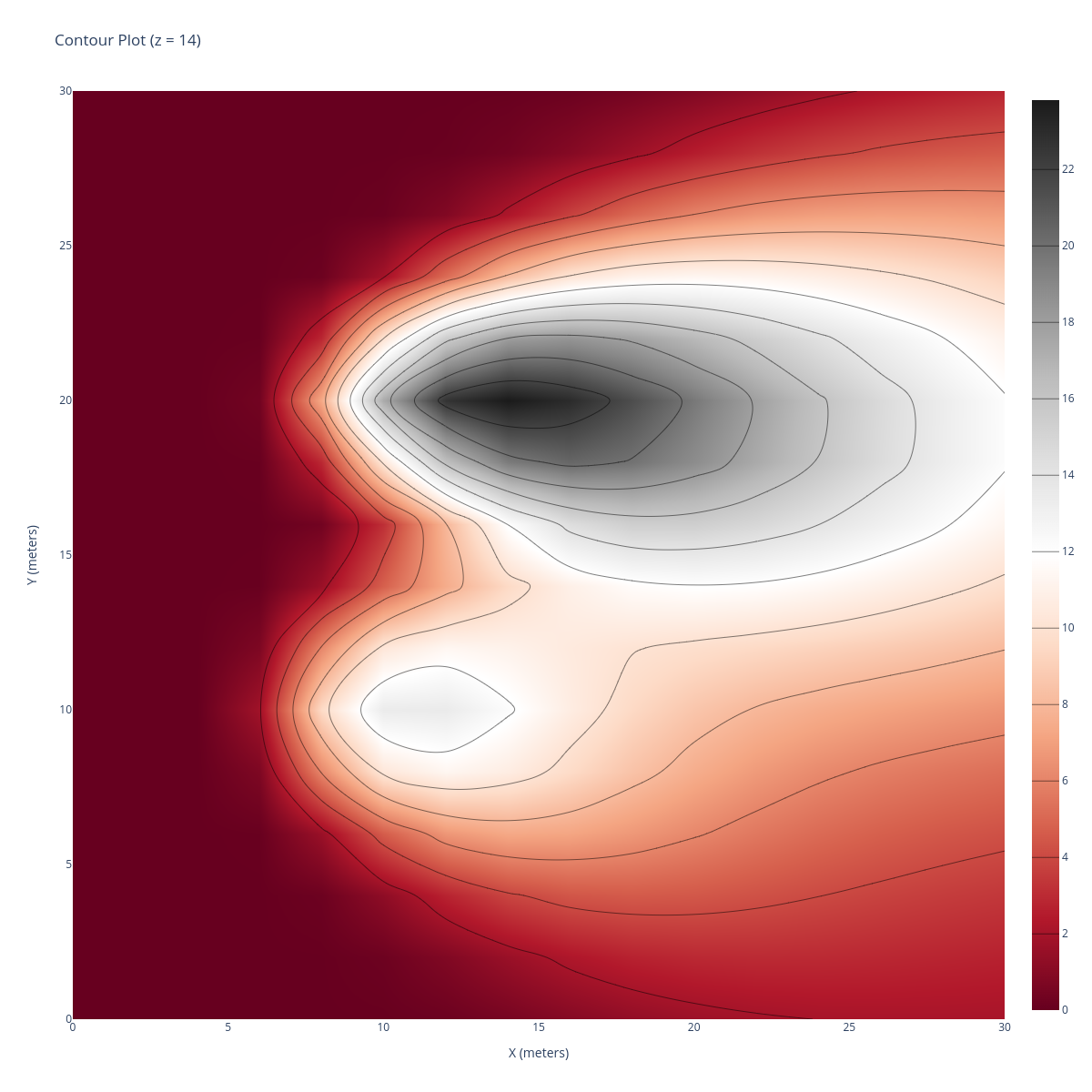}
        \label{fig:Contour-plot}
    }
    \caption{Plume dispersion simulations using Eq. \ref{eq:multiple-point-source} with two point source emitters, $C_1$ and $C_2$, characterized by distinct wind speeds ($u$), emission rates ($Q$) and effective heights ($H$). For $C_1$, $u = 10$ m/s, $Q=10$ g/s, $H=16$ m, under stability class A conditions; for $C_2$, $u = 8$ m/s, $Q=12$ g/s, $H=16$ m, under stability class B conditions. These meteorological parameters influence the plume behavior and dispersion patterns observed.}
    \label{fig:plume_example}
\end{figure}

\noindent where C is the pollution concentration, K is the diffusivity tensor, u is the time-averaged wind velocity and S is a source function. In steady-state conditions $ \frac{\partial C}{ \partial t} = 0 $, the pollutant is emitted at constant rate $ Q $ [kg/s] from a continuous point source located at $x = (0, 0, H)$, where $H$ is referred to as the stack height. We assume advection via an uni-directional wind oriented along the x-axis $\vec{u} = (u, 0, 0)$ for some constant $ u \geq 0$. If wind velocity is sufficiently large, then the contribution of downwind diffusion becomes negligible, thus $K_x = 0$. Lastly, we assume eddy diffusivity is isotropic and is a function of the downwind distance only, so that $K_y(x) = K_z(x)$. Under the aforementioned assumptions, Equation \ref{eq:advection-difussion equation} is simplified to the following second-order PDE:

\begin{equation} \label{eq:gpm-pde}
    u \frac{\partial C}{\partial x} = K \frac{\partial^2 C}{\partial y^2} + \frac{\partial^2 C}{\partial z^2} + Q\delta(x) \delta(y) \delta(z-H)
\end{equation}

\noindent where $\delta$ is the Dirac delta with boundary conditions: 

\begin{equation} \label{eq:boundary-conditions}
    \begin{aligned} 
            C(0, y, z) = 0, \ C(\infty, y, z) = 0, \ C(x, \pm \infty, z) = 0, 
            \\ \ C(x, y, \infty) = 0, \  K(\frac{\partial C}{\partial z}(x, y, 0) = 0
    \end{aligned}
\end{equation}

The initial condition reflects the unidirectional wind flow assumption, such that no contaminant is found at $x < 0$. Subsequent infinite-boundary conditions are set to ensure that the overall mass of the contaminant remains finite, in alignment with mass conservation principles. The last condition stipulates the vertical flux at ground level is zero, a condition that follows from the assumption that the ground is neither a source nor sink for the pollutant.

Equations \ref{eq:gpm-pde} and \ref{eq:boundary-conditions} define a well-posed problem for steady-state pollution concentrations. Laplace transforms and Green's functions are two of the several existing techniques that can be leveraged to solve our PDE, for which complete derivations can be found in  \citep{doi:10.1137/10080991X}, \citep{DerivGaussianVeigele}. Thus, the closed-form analytical solution is described by: 

\begin{equation} \label{eq:gpm}
    \begin{aligned} 
    C(x, y, z) = \frac{Q}{2\pi u\sigma_y\sigma_z} \exp\left[ - \left(\frac{y^2}{2 \sigma_y^2} \right) \right] \times \\ 
    \left[ \exp\left[ \frac{-(z - H)^2}{2\sigma^2_z}\right] + \exp\left[ \frac{-(z + H)^2}{2\sigma^2_z}\right] \right]
\end{aligned}
\end{equation}
where $\sigma_y$ and $\sigma_z$ represent the respective lateral and vertical spreads normal to wind direction. $u$. 

Equation \ref{eq:gpm} can be further extended to model the coalesced pollution plume from multiple constituent point sources. Let $n$ denote the total number of point sources, where $\vec{p_i} = (X_i, Y_i, H_i)$ is the location of the $i$th pollutant source with $Q_i$ emission rate. Then the contaminant concentration at any point $(x,y,z)$ is simply the superposition of their individual sources \citep{doi:10.1137/10080991X}

\begin{equation}\label{eq:multiple-point-source}
    C_T(x,y,z) = \sum_{s=1}^{n} C(x^{\prime}_i, y^{\prime}_i, z; Q_i, H_i)
\end{equation}

\noindent where $x^{\prime}_i$ and $y^{\prime}_i$ are the shifted coordinates such that $x^{\prime}_i = y^{\prime}_i = 0$ for each $i$. 

\begin{equation*}
    x^{\prime}_i = x - X_i \ \text{and} \ y^{\prime}_i = y - Y_i
\end{equation*}

To train our multi-agent reinforcement models, we leveraged Equation \ref{eq:multiple-point-source} to create multiple instances of pollution maps with varying meteorological conditions, number of point sources and emission rates. 
Fig. \ref{fig:plume_example} displays a sample plume environment with two pollution sources along with its corresponding contour plot at a fixed height.

\section{Modeling}\label{sec:modeling}
In the following sections, we delve into two concepts pertinent to our research: Partially Observable Markov Decision Games (POMGs) and Recurrent Reinforcement Learning (RRL) techniques. POMGs extend the Markov Decision Process framework to multi-agent systems with incomplete information. RRL techniques, particularly Deep Recurrent Q-Networks (DRQN) and Double Deep Recurrent Q-Networks (DDRQN), utilize LSTM units to manage temporal dependencies in partially observable settings. These models serve as baseline algorithms in our experiments, against which we benchmark the performance of our proposed ADDRQN algorithm. 

\begin{figure}[!htb]
    \centering
    \includegraphics[width=.95\columnwidth]{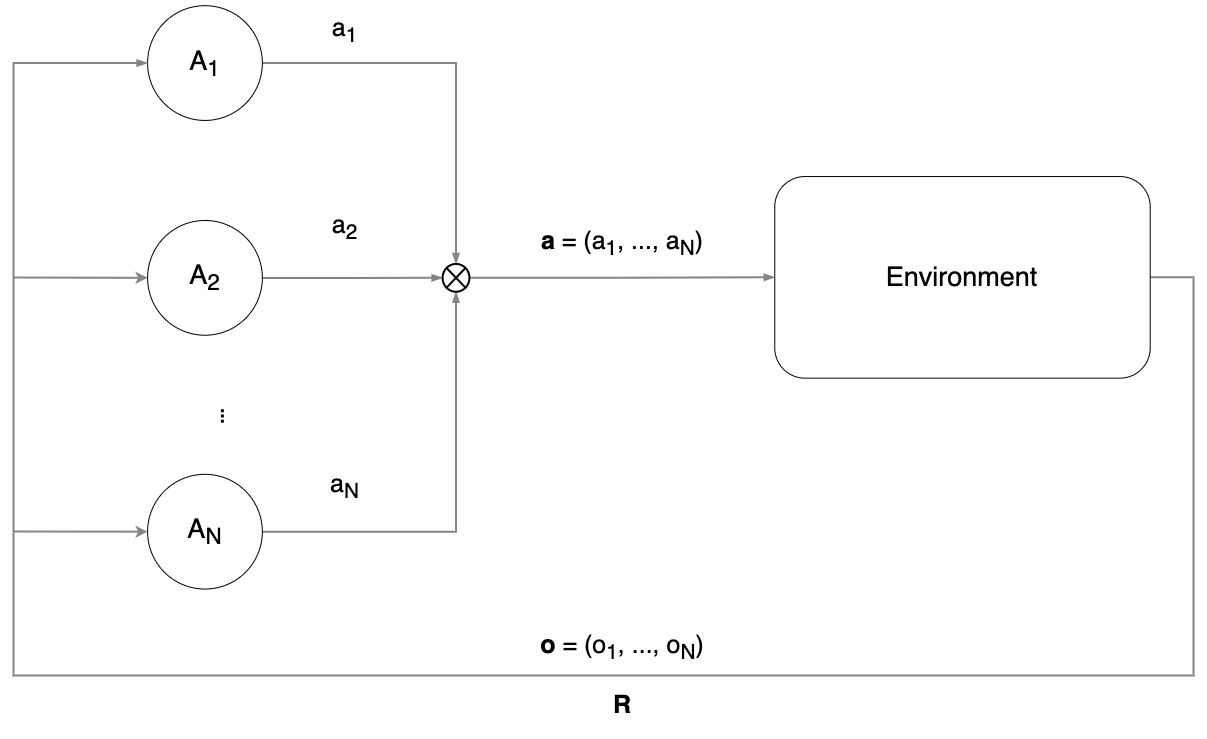}
    \caption{Structure of a Partially Observable Markov Game (POMG). Individual decisions from agents contribute to a joint action vector $\mathbf{a}$ which influences the shared environment. This results in a joint observation vector and an immediate global reward for the agents.}
    \label{fig:POGM}
\end{figure}

\subsection{Multi-Agent Reinforcement Learning}
Multi-Agent Reinforcement Learning (MARL) extends the principles of Reinforcement Learning (RL) to scenarios involving multiple agents that interact within a shared environment. These interactions can be cooperative, competitive, or a mixture of both, depending on the task requirements and the nature of agent interactions. In cooperative settings, agents work together to achieve a common goal, whereas in competitive settings, agents have conflicting objectives. Mixed settings involve a combination of cooperation and competition among agents.

MARL training regimes can be categorized into Independent Learners (IL), Joint Action Learners (JALs), and Centralized Training with Decentralized Execution (CTDE) \citep{zhang2021multiagentreinforcementlearningselective}. Independent Learners (IL) train each agent separately, treating other agents as part of the environment. This approach, while simple and scalable, often struggles with non-stationarity due to the evolving policies of other agents. JALs learn a joint Q-function over the combined action space of all agents, ensuring better coordination and stability but often criticized for computational challenges due to the exponential growth of the joint action space. The CTDE approach leverages centralized training with access to global state information and decentralized execution, where agents operate based on local observations, balancing coordinated learning with practical execution. Our work adopts the JAL framework in a fully cooperative environment, in which multiple UAVs collaborate to trace and localize airborne pollutants. The JAL framework is particularly suited for our case as it allows for the precise coordination required in plume tracing tasks. By considering the joint actions and states of all UAVs, our approach ensures that the agents work together effectively to optimize the detection and localization process. While CTDE offers a balanced approach, the necessity for high-level coordination and the fully cooperative nature of our task makes JAL a more fitting choice. The ability of JAL to provide stable and coordinated learning outcomes, despite its computational demands, aligns well with the complex and dynamic nature of the environment we are addressing. This leads us to our adoption of Partially Observable Markov Decision Games (POMGs), a suitable framework for modeling our multi-agent system under partial observability.

\subsection{Partially Observable Markov Decision Games}
\label{sec:POMG}

MDPs provide a framework for decision-making in fully observable, single-agent environments, where agents aim to maximize cumulative rewards through a series of state-action transitions \citep{PUTERMAN1990331}. Extending MDPs, Partially Observable Markov Decision Process (POMDP) adapt to scenarios where an agent may have incomplete information about the state, using observation functions and belief states to approximate the environment’s true state \citep{annurev:/content/journals/10.1146/annurev-control-042920-092451}. In this work, we adopt the formalism introduced in Partially Observable Markov/Stochastic Games (POMGs), which extends the POMDPs framework to multi-agent systems.

POMG are formally described by an 8-tuple \( \langle N, S, \{A_i\}, P, R, \gamma, \{O_i\}, Z \rangle \), where \(N\) represents the number of agents interacting within the environment, \(S\) is the set of all possible states that the environment may assume, and \(\{A_i\}\) indicates the specific set of actions available to each agent \(i\). The function \(P : S \times \mathcal{A} \times S \rightarrow [0, 1]\) defines the state transition probabilities, where \(\mathcal{A} = A_1 \times \ldots \times A_N\) represents the set of all possible joint actions. The reward function \(R : S \times \mathcal{A} \times S \rightarrow \mathbb{R}\) determines the rewards received based on the states and actions, and \(\gamma\) serves as the discount factor influencing the valuation of future rewards. Each agent has access to a distinct set of observations, represented by \(\{O_i\}\), which collectively form the joint observation set \(\mathcal{O} = O_1 \times \ldots \times O_N\). The observation function \(Z : S \times \mathcal{A} \times \mathcal{O} \rightarrow [0, 1]\) maps the state and joint actions to a probability distribution over these joint observations, enabling agents to make informed decisions despite partial observability.

In a POMG, agents perceive the environment through these limited, partial observations \( o \in O \), influenced by the observation function \(Z\), which calculates the likelihood of observing \( \mathcal{O} \) based on the collective action \( \mathcal{A} \) and resulting state. To navigate this uncertainty, agents may adopt strategies such as maintaining an observation history \( H_t = \{(o_1, a_1), \ldots, (o_{t-1}, a_{t-1})\} \) or developing belief states that involve updating a probability distribution over possible states with each new piece of data. This work utilizes a historical approach, implemented through LSTM networks, which process action-observation history to distill it into a high-level internal state, capturing long-range dependencies and key events that inform subsequent action choices in a partially observable setting.

Agents interact with the environment and each other through a series of discrete time steps. At each time step \(t\), each agent observes part of the environment's state through its observations \(o_i \in O_i\), based on which it selects an action \(a_i \in A_i\). These actions form a joint action vector \( \mathbf{a} = (a_1, \ldots, a_N)\), and is executed on the environment, leading to a transition to a new state \(s'\) according to the transition probability \(P(s' | s, \mathbf{a})\). The agents then receive a joint observation \( \mathbf{o} = (o_1, \ldots, o_N)\) reflecting the new state \(s'\), determined by the observation function \(Z\), and a collective immediate reward determined by \(R\).  This interactive loop is depicted in Figure \ref{fig:POGM}. The objective of the agent is to learn an optimal policy $\pi(a|s)$, a mapping from environment states to actions that maximizes the long-term cumulative reward function:

$$ R_t = r_t + \gamma r_{t+1} + \gamma^2 r_{t+2} + ... = \sum^{\infty}_{k=0} \gamma^k R_{t+k+1}$$

\subsection{Recurrent Reinforcement Learning}\label{sec:Deep Reinforcement Learning}

Deep Recurrent Q-Networks (DRQN)  and Double Deep Recurrent Q-Networks (DDRQN) build upon the foundations of Q-learning to address the complexities of environments characterized by POMDPs. These models aim to compute a RNN-based Q-function parameterized by $\theta$, $Q(o, a | \theta, h)$ representing the expected cumulative discounted future reward for taking an action $a$ in an observed state $o$ under a given policy $\pi$, where $h$ denotes the LSTM hidden state. The key advancement in DRQN and DDRQN is the integration of Recurrent Neural Networks (RNNs), specifically LSTM units, which allow these models to capture temporal dependencies and nuances between successive observations. Within these LSTM-equipped models, the hidden state \(h_t\) is dynamically updated at each timestep \(t\) based on the current observation \(o_t\) and the preceding hidden state \(h_{t-1}\), expressed as \( h_t = \text{LSTM}(o_t, h_{t-1})\). This updated hidden state then feeds into a fully connected layer that outputs the Q-values, using \(h_t\) as input, thereby effectively integrating historical context into each decision point. This enhancement is particularly effective in bridging the gap between observed states $o$ and true states $s$, improving the model's accuracy by approximating that $Q(o, a | \theta) \approx Q(s, a | \theta)$.

The network parameters $\theta$ are optimized iteratively through the minimization of a differentiable loss function, calculated on mini-batches of state transitions $(o, a, r, o')$ that are uniformly sampled $U(D)$ from a replay buffer $D$:
\begin{equation}\label{eq:loss-function}
    L(\theta) = \mathbb{E}_{(o, a, r, o') \sim U(D)} \left[ \left( y - Q(o, a; \theta; h) \right)^2 \right]
\end{equation}

\noindent In the DRQN framework, the target value $y$  is defined as:

\begin{equation}
     y =   r + \gamma \max_{a'} Q(o', a'; \theta^-; h') 
\end{equation}

\noindent where $\theta^-$ denotes the parameters of a target network, a periodically updated copy of the online network that provides more stable Q-value targets. The target network is a key stabilization component that decouples the next-state and current Q-value functions. This helps avert learning oscillations and divergence issues outlined in \citep{580874}. The use of a replay buffer, which stores recent transitions and allows for random sampling of mini-batches for updates, breaks the correlations introduced in sequential sampling and reduces variance in updates.

Building on DRQN, Double Deep Q-Network (DDRQN) addresses the overestimation bias inherent in the max operator used by DRQN by decoupling action selection from evaluation \citep{vanhasselt2015deep}. DDRQN  uses online network to select the optimal action and a target network to evaluate its Q-value. The target value $y$ becomes: 

\begin{equation}\label{eq:ddrqn-loss}
    y = r + \gamma Q(o', \arg \max_{a'} Q(o', a'; \theta; h'); \theta^-; h' )
\end{equation}

\noindent This approach reinforces the importance of observation states $o$ as proxies for the agent’s perceivable environment, in accordance with the principles of POMDP/POGM frameworks.

\begin{algorithm}[!htb]
\begingroup
\fontsize{10pt}{10pt}\selectfont
\DontPrintSemicolon
\caption{Multi-Agent Reinforcement Learning Pipeline with ADDRQNs}\label{alg:multi_agent_cap}
Initialize Experience Replays \( D^i \) for each agent \( i \)\;
Initialize hidden states \( h^i \) and actions \( a^i \) for each agent \( i \)\;
Initialize environment, Agents, \( \epsilon \), and other constants\;
\For{a finite number of Episodes}{
    Choose pollution environment, reset state to some initial state \( s \)\;
    Add noise  \(\nu_t \sim \mathcal{N}(0, \sigma^2) \) to each pollution state in plume environment, where \( \sigma = k \cdot |C_T(p_t)| \) is proportional to the raw magnitude}\;
    Reset hidden states \( h^i \) and last\_actions \( a^i \) for each agent \( i \)\;
    \For{a finite number of Epochs}{
        Choose actions \( a^i \) given observed state \( o_t \) using \( \epsilon \)-greedy policy for each agent \( i \)\;
        Agents execute joint action \( \mathbf{a}_t \), observe a global immediate reward \( \mathbf{r}_t \), and next state \( \mathbf{o}_{t+1} \)\;
        Store transition \( (a_{t-1}^i, o_t^i, a_t^i ,r_t^i, o_{t+1}^i, \text{done}^i) \) in experience replay buffer \( D^i \) for each agent \( i \)\;
        Update observation \( \mathbf{o}_t \leftarrow \mathbf{o}_{t+1} \)\;
        Update prior action \( \mathbf{a}_{t-1} \leftarrow \mathbf{a}_{t} \) for each agent $i$\;
        \If{episode \( > \) explore}{
            \( \epsilon \leftarrow \) \( \epsilon_{\text{new}} \) with \( \epsilon \)-decay using Eq. \ref{eq:e-decay}\;
            \For{each agent \( i \)}{
                Uniformly sample a mini-batch of N sequential transitions from \( D^i \)\;
                Compute batched target $\hat{Q}^i$-values using \( \mathbf{y}^i = \mathbf{r}^i + \gamma \hat{\mathbf{Q}}^i(\mathbf{o}_{t+1}, \text{arg max} \, \mathbf{Q}^i(\mathbf{o}_{t+1}, \mathbf{a}_{t+1})) \)\;
                Calculate the loss using \( L^i = \frac{1}{N} \sum_{t=1}^{N} \left( Q^i(o_t, a_t) - \mathbf{y}^i_t \right)^2 \)\;
                Update online network parameters using back propagation and optimizer\;
                Update target network parameters using soft update \( \theta^-_i = \tau \theta_i + (1 - \tau) \theta^-_i \)\;
            }
        }
        \If{done}{
            break\;
        }
    }
\endgroup
\end{algorithm}

\begin{figure}[!h]\label{fig:addrqn-arch}
    \includegraphics[width=\columnwidth]{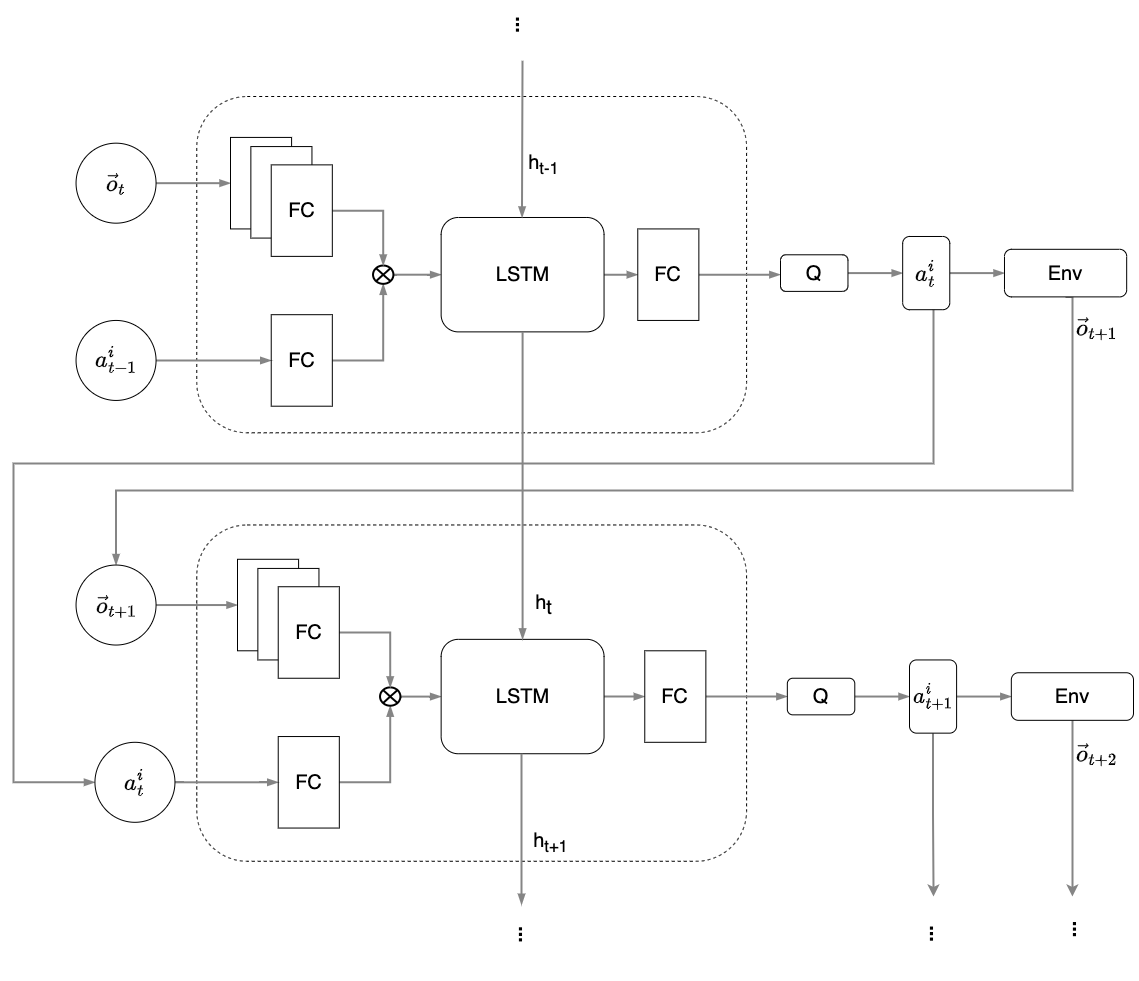}
    \caption{Recurrent Q-network architecture over two sequential time steps. The network integrates fully connected layers with an LSTM unit to process observations ($\vec{o}_t$) and preceding actions ($a^i_{t-1}$). This allows for estimating Q-values and choosing subsequent actions while maintaining temporal dependencies across episodes. Architecture adapted from \citep{zhu2018improving}. }
    
\end{figure}

\section{MARL Framework}
\label{sec:ADDRQN}

\begin{figure}[!h] \label{fig:marl-arch}
    \includegraphics[width=\columnwidth]{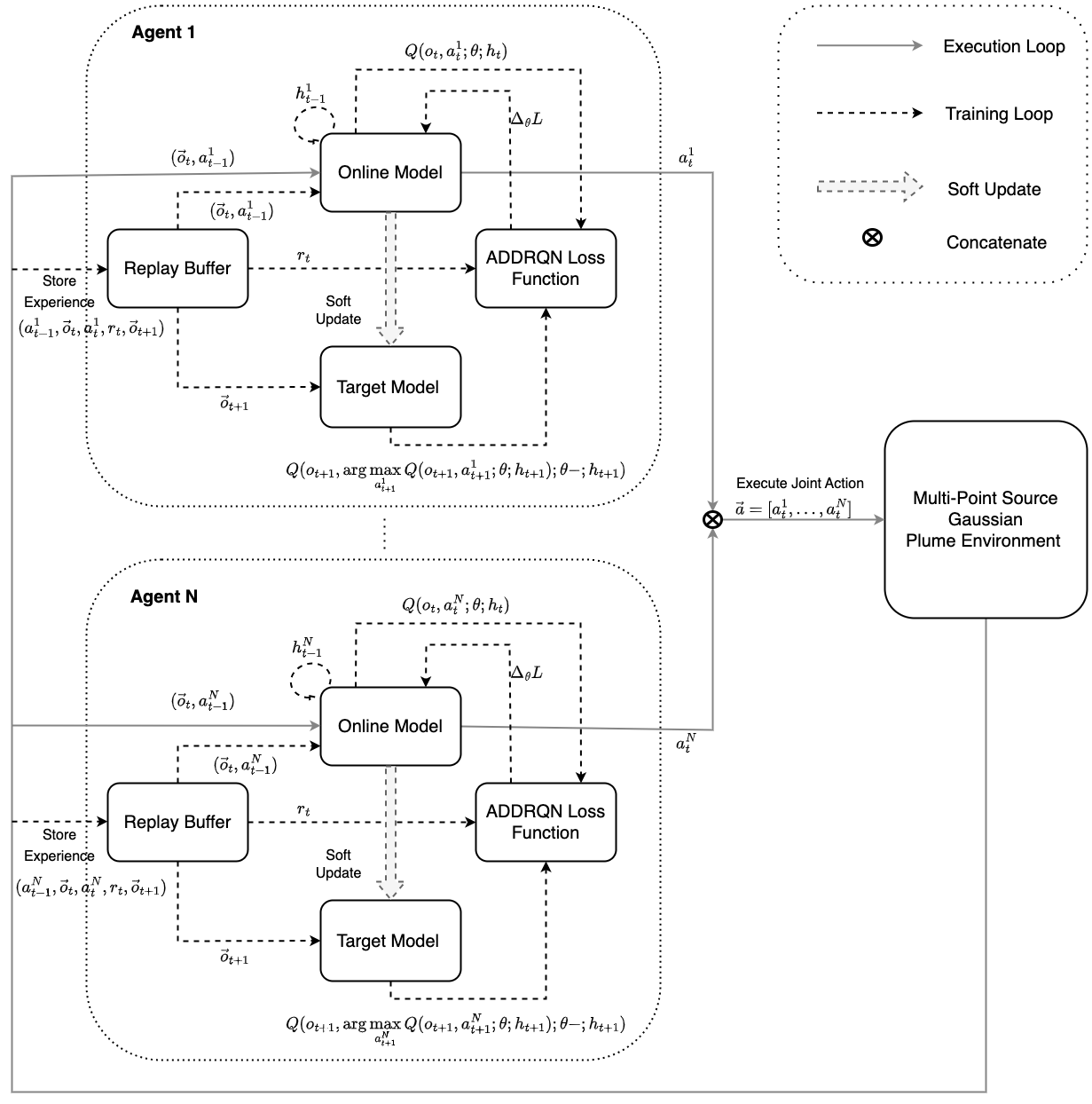}
    \caption{ }
\end{figure}

The presented architecture details a Multi-Agent Reinforcement Learning (MARL) framework developed for tracing multiple airborne pollution sources using sUAS in a Gaussian plume environment. The overall framework amalgamates techniques detailed in subsections \ref{sec:modeling}. Within this framework, we introduce the novel Action-Specific Double Deep Recurrent Q-Network (ADDRQN), which judiciously incorporates joint observations and prior actions as inputs into the Recurrent Q-network, inspired by empirical findings from Zhu et al., who demonstrated improved decision-making in partially observable settings by employing observation-action pairs in LSTM inputs \citep{zhu2018improving}. This framework operates through two primary loops: the Execution Loop and the Training Loop, as depicted in Figure \ref{fig:marl-arch}.

In the Execution Loop, each UAV independently acquires environmental data, computes decisions, and executes actions. At each timestep $t$, each agent $i$ receives a joint observation vector $\vec{o} = (o_1,...,o_n)$ that encapsulates aggregated sensory data across the UAV network, alongside a one-hot encoded action $a_{t-1}^i$ from the preceding timestep. This data is processed by the Online Model, specifically an ADDRQN neural network $Q(o,a ; \theta; h)$ featuring LSTM units and densely connected layers as illustrated in Figure \ref{fig:addrqn-arch}. Our ADDRQN model processes joint observations and actions $(\vec{o}, a_{t-1}^i)$, generates embeddings for the observations and actions, concatenates them, processes them through LSTM units, and outputs Q-values corresponding to every possible action. Each agent locally selects the optimal action $a_t^i$ based on the highest Q-value, aggregates these individual actions into a joint action vector $\vec{a} = (a^1_t, ..., a^N_t)$, and collectively executes this vector within the simulated plume environment. This process yields an immediate global shared reward $r_t$ and each agent observes a new observation state $o_{t+1}$. These elements constitute an experience tuple $({a_{t-1}, o_t}, a_t, r_t, o_{t+1})$ that is stored in an Experience Replay Buffer. The loop continues until a terminal condition is met: either when all pollution plumes are found or after exceeding a predetermined number of timesteps. 

Within the Training Loop, the primary objective is to update the Online and Target neural nets to refine future decision-making. To facilitate this process, each agent stores its experiences as tuples $(\{a_{t-1}, o_t\}, a_t, r_t, o_{t+1})$ in the Experience Replay Buffer. During training, mini-batches are uniformly sampled from this buffer to calculate the difference between the predicted Q-values from the Online Model and the target Q-values from the Target Model using Equations \ref{eq:ddrqn-loss} and \ref{eq:loss-function}. The Target Model, an intermittently updated copy of the Online Model, provides stable Q-value targets to prevent training divergence. Optimization is carried out using the Adam optimizer, and the loss function  is minimized accordingly. Parameters for the primary network are updated during the backward pass, while the target network's parameters are softly updated. The soft update is given by

\begin{equation}
    \theta^- = \tau * \theta + (1 - \tau) * \theta^-
\end{equation}

\noindent where $\theta$ and $\theta^-$ are the online and target network parameters, respectively, and $\tau$ is a hyperparameter that controls the degree of the soft update, where $\tau \ll 1$. Our framework employs an exponentially decaying $\epsilon$-greedy strategy to balance the exploration-exploitation trade-off. Initially, we designate a fixed number of exploration episodes, during which actions are randomly sampled from the action space ($\epsilon = 1$). As training unfolds, we progressively decay $\epsilon$ according to the following formula: 

\begin{equation} \label{eq:e-decay}
    \epsilon(\text{t}) = \epsilon_{\text{min}} + (\epsilon_{\text{max}} - \epsilon_{\text{min}}) \times \exp\left(-\frac{t - T_{\text{init}}}{\lambda}\right)
\end{equation}

\noindent Here, the decay commences after a predefined episode count, $T_{\text{init}}$, and asymptotically approaches $\epsilon_{min}$, at a decay rate controlled by the decay constant $\lambda$. During the exploitation phase, each agent updates the neural network's weights using uniformly sampled mini-batches from the Experience Replay buffer. The pseudo-code detailing our proposed multi-agent algorithm is outlined in Algorithm \ref{alg:multi_agent_cap}.

\section{Experimental Design}
\label{sec:Exp_design}

Our MARL algorithms were trained on a series of synthesized Gaussian global plumes, each represented as $C_T(x, y, z)$, constructed by varying meteorological conditions and pollution source counts. Specifically, we systematically generated a list of pollution sources by iterating through specified ranges of wind speed, emission rate, effective height, and stability parameters. Each unique parameter combination yields a distinct plume instance denoted by $C(x,y,z)$, centered at the origin and whose pollution values are governed by Equation \ref{eq:gpm}. Thereafter, two individual plumes were randomly selected from this list, and their coordinates were shifted to random positions $L_j$ on the map. These sources were combined to synthesize a global plume $C_T(x, y, z)$ via the superposition of individual pollution sources as described by Eq. \ref{eq:multiple-point-source}. Our MARL algorithms were trained and tested on 40 unique synthesized global plumes $C_T(x, y, z)$ derived from the aforementioned procedure. The dataset was split 50/50, with 20 plumes used for training and 20 for testing.

\begin{table}[!htb]
  \centering
  \caption{Hyperparameters for MARL algorithm}
  \begin{tabularx}{\columnwidth}{Xc} 
    \toprule
    Hyperparameter & Value \\
    \midrule
    Learning Rate ($\alpha$) & 0.001 \\
    Tau ($\tau$) & 0.01 \\
    Batch Size & 768 \\
    Sequence Length & 75 \\
    Hidden Size & 1024 \\
    Max Epochs & 5000 \\
    Gamma ($\gamma$) & 0.999 \\
    Epsilon Start ($\epsilon_{max}$)& 1.0 \\
    Epsilon End ($\epsilon_{min}$)  & 0.05 \\
    Decay Rate ($\lambda$)  & 100 \\
    Exploration Episodes  ($T_{init}$) & 22000 \\
    Total Episodes & 23000 \\
    \bottomrule
  \end{tabularx}
  \label{tab:marl_hyperparameters}
\end{table}

\begin{figure*}
    \centering
    \vspace{-15pt}
    \subfigure[]{
        \includegraphics[width=.45\textwidth]{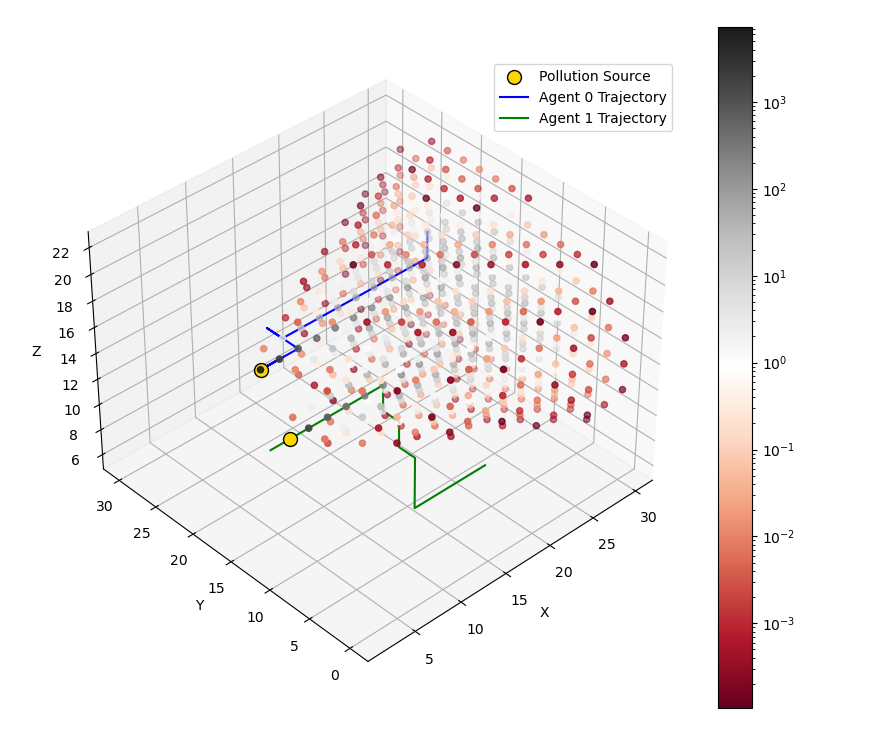}
        \label{fig:trajectory-1}
    }
    \hfill 
    \subfigure[]{
        \includegraphics[width=.45\textwidth]{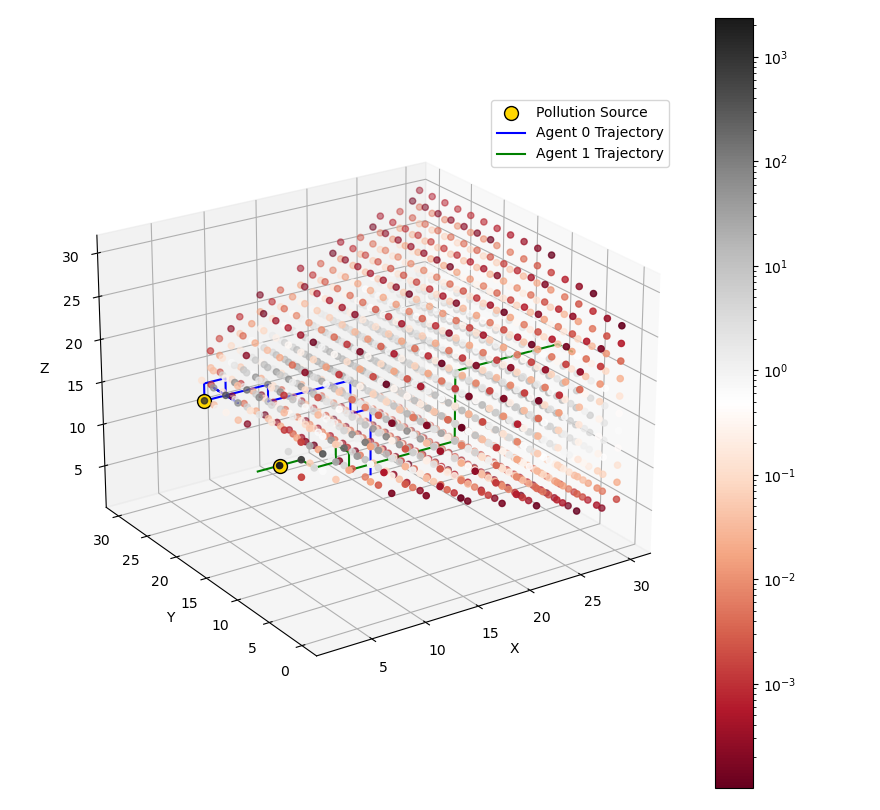}
        \label{fig:trajectory-2}
    }
    \caption{Agent trajectories under varying plume dispersion conditions with adjacent color bar indicating corresponding pollution level.}
    \label{fig:addrqn_trajectory}
\end{figure*}

During each training episode, a global plume was uniformly sampled, where each agent's starting position was chosen at random. To improve generalizability of our models, we added noise to each pollution state at each episode. Specifically, for each pollution observation $C_T(p_t)$, we added a noise $\nu_t$ drawn from a normal distribution,  $\nu_t \sim \mathcal{N}(0, \sigma^2)$, where the standard deviation $\sigma = k \cdot |C_T(p_t)|$ is proportional to the raw observed pollution magnitude, with  $k$ set to vary between 5\% and 10\%. This range was chosen to mimic the variability encountered in real-world sensor data, where measurement accuracy can be affected by factors such as sensor calibration, environmental conditions, and the inherent instability of the pollutant dispersion. At each epoch $t$, the  $i$th UAV observes its local geolocation $p_t = (x,y,z)$ and the corresponding pollution concentration $C_T(p_t)$. The collected geolocation and pollution concentrations of all agents form a joint observation vector $\textbf{o}_t$, used by the Q-network $Q(o_t, a_t; \theta; h_t)$ to ascertain the optimal action. Correspondingly, each agent selects an action $a_i$ from the set $ \mathcal{A}\in [\text{Left, Right, Up, Down, Forward, Backward}]$; the joint action is executed in the shared environment and agents are given an immediate global reward based on the resulting observational state. We define the reward function as:

\begin{equation}
    r = \begin{cases}
    10 & \text{if } p_i = L_j \text{ and } L_j \notin V\\
    0 & \text{if } p_i = L_j \text{ and } L_j \in V\\
    -1 & \text{otherwise}
    \end{cases}
\end{equation}

\noindent where $V$ denotes a set of pollution sources previously located by an agent. The agent's actions are evaluated based on the resulting observation states, receiving a penalty of -1 for each time step that an agent fails to find a pollution source and a reward of 10 for discovering a new source. Agents receive no reward for revisiting a previously localized source. This reward structure was designed to incentivize efficient pollution source discovery. Training episodes conclude either once all pollution origins have been located or when the number of epochs exceed a predefined threshold. Each algorithm was trained using the hyperparameters listed in Table \ref{tab:marl_hyperparameters}.

\section{Performance}\label{sec:Performance}

This section presents a comprehensive analysis of our proposed ADDRQN algorithm's performance. We conduct two main sets of experiments: a comparative analysis against state-of-the-art baseline algorithms and multi-agent experiments to investigate the impact of varying agent numbers on performance.

For both sets of experiments, we conducted evaluations consisting of 10,000 episodes. Maps and starting positions uniformly sampled and held constant across all models to maintain evaluation parity. To ensure a fair and unbiased comparison, all learning-based algorithms were subjected to identical training conditions and hyperparameters, as detailed in Table \ref{tab:marl_hyperparameters}. Given the stochastic nature of the learning process,  we trained five independent models for each algorithm to account for variability in performance. 

 To quantify and compare the algorithms' performance, we employed four key performance indicators (KPIs): \textbf{Success Rate}, which represents the percentage of trials in which all pollution sources were successfully located within the predefined number of epochs -- where a higher success rate indicates superior effectiveness in achieving the primary objective; \textbf{Average Epochs}, an efficiency metric reflecting the average number of epochs required to locate all pollution sources per episode -- where lower average epochs signify faster task completion;  \textbf{Total Rewards Per Episode}, which suns the rewards accumulated per epoch, with higher total rewards per episode suggesting more optimal path planning and decision-making capabilities; and \textbf{Training Time}, which measures the computational efficiency of each model during the learning phase. The highest-performing model from each algorithmic category was then chosen for subsequent analysis.

\subsection{Comparative Analysis}

\begin{table*}[!htb]
  \centering
  \small
  \caption{Performance summary of RL models}
  \begin{tabular}{lccccc}
    \toprule
    Metric & ADDRQN & ADRQN & DDRQN & DRQN &   Random\\
    \midrule
    Success Rate    & 95.97\%   & \textbf{98.66\%}   & 74.21\%   & 62.69\% &  8.0 \%\\
    Avg Epochs      & \textbf{52.07}     & 59.12     & 86.92     & 84.80   &   618.41\\
    Avg Reward      & \textbf{-28.07}    & -35.12    & -62.92   & -60.80   &   -594.41\\
    Training Time   & \textbf{3.16}      & 4.02      & 6.83   &  5.35    &    -- \\
    \bottomrule
  \end{tabular} 
  \label{tab:rl_model_performance}

\end{table*}

In this study, we conducted a comprehensive comparative analysis to assess the performance of our proposed ADDRQN algorithm against three state-of-the-art baseline algorithms: DRQN, DDRQN, and ADRQN, as well as a random walk baseline that performs random exploration of the environment. This random baseline served as a stochastic benchmark to contextualize the aptitude of learning-based methods.

\begin{figure*}[!htb]
    \centering
    \subfigure[ADDRQN]{
        \includegraphics[width=.48\textwidth]{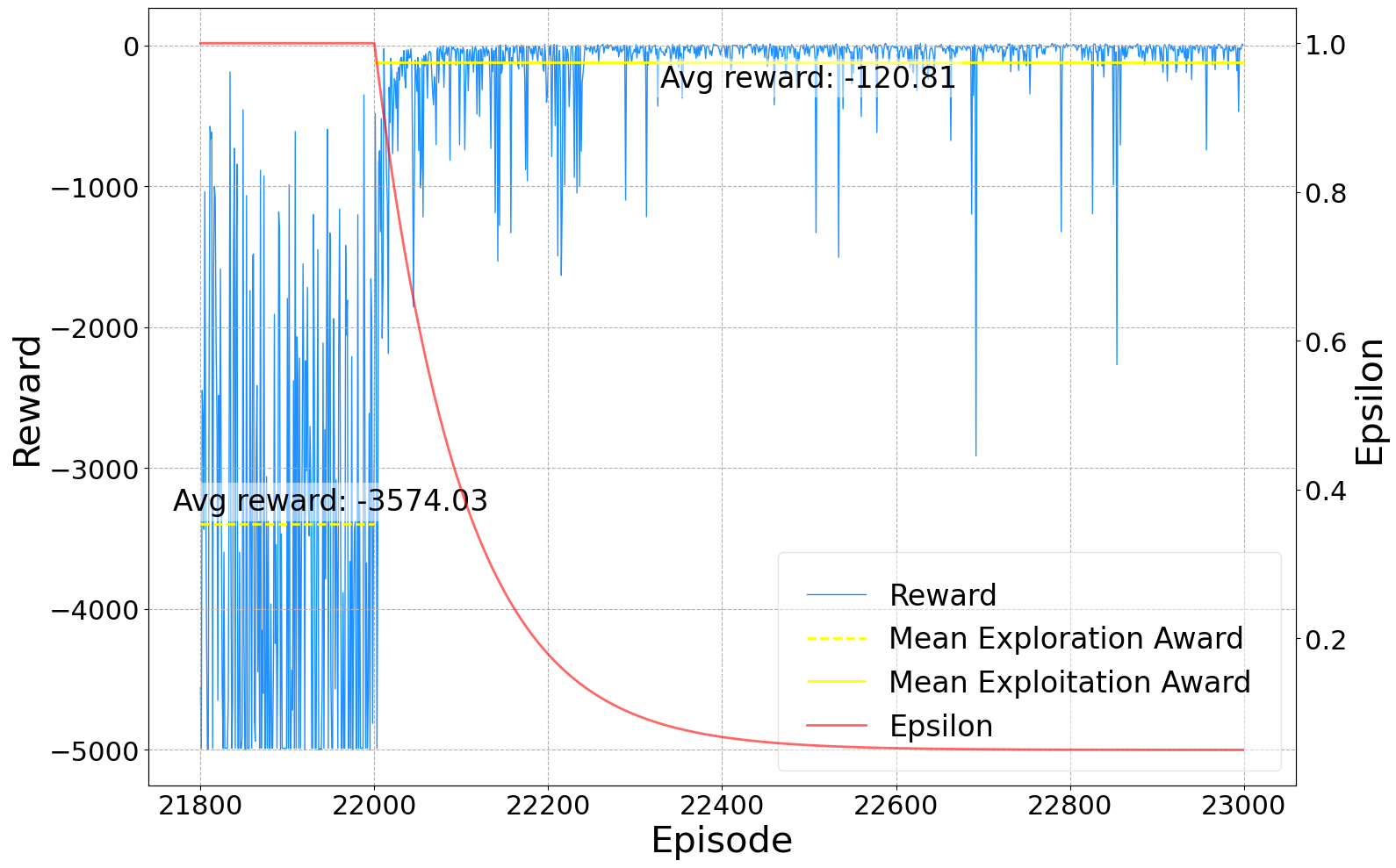}
        \label{fig:award-addrqn-subset}
    }
    \hfill
    \subfigure[ADRQN]{
        \includegraphics[width=.48\textwidth]{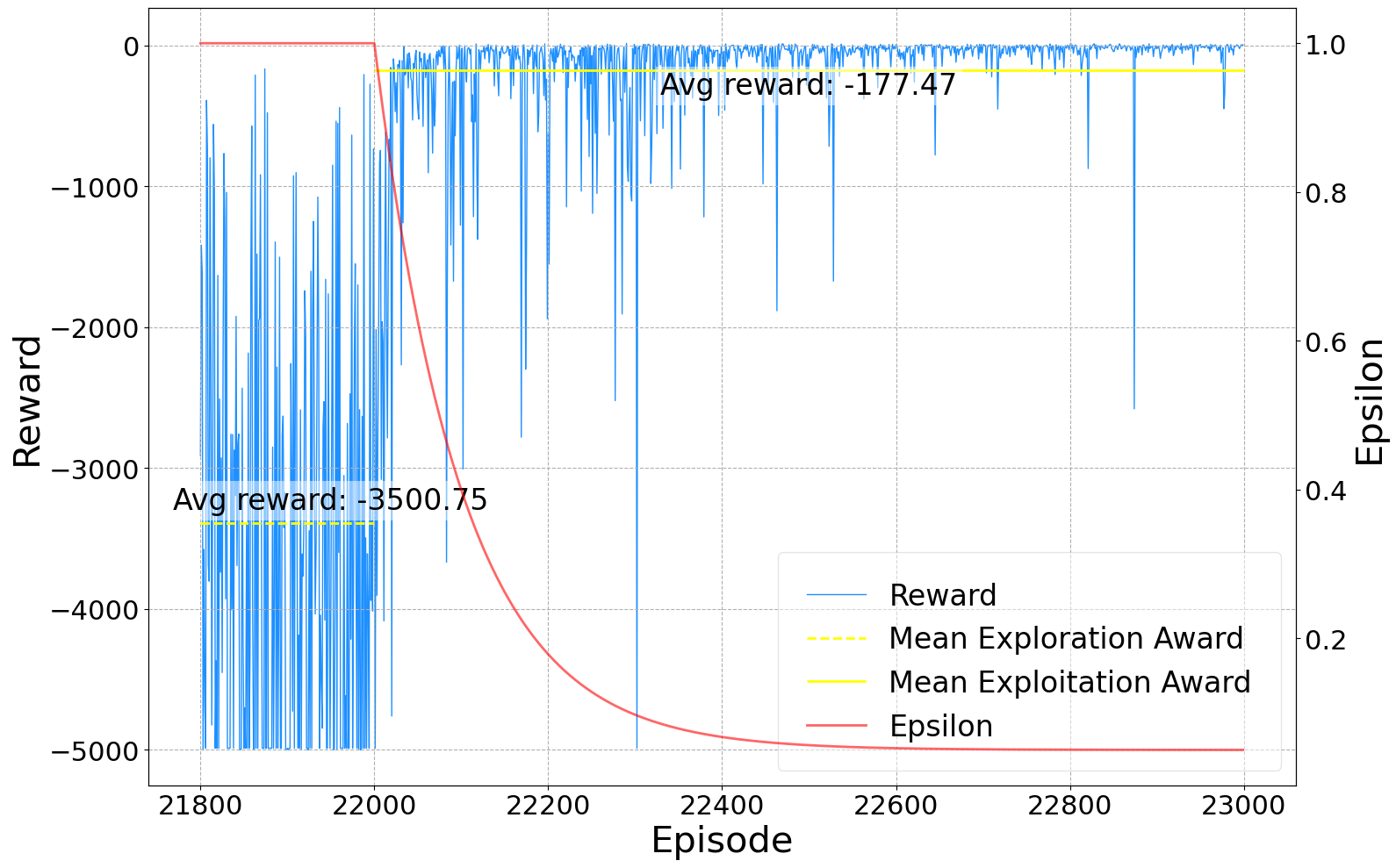}
        \label{fig:award-adrqn-subset}
    }
    \\
    \subfigure[DDRQN]{
        \includegraphics[width=.48\textwidth]{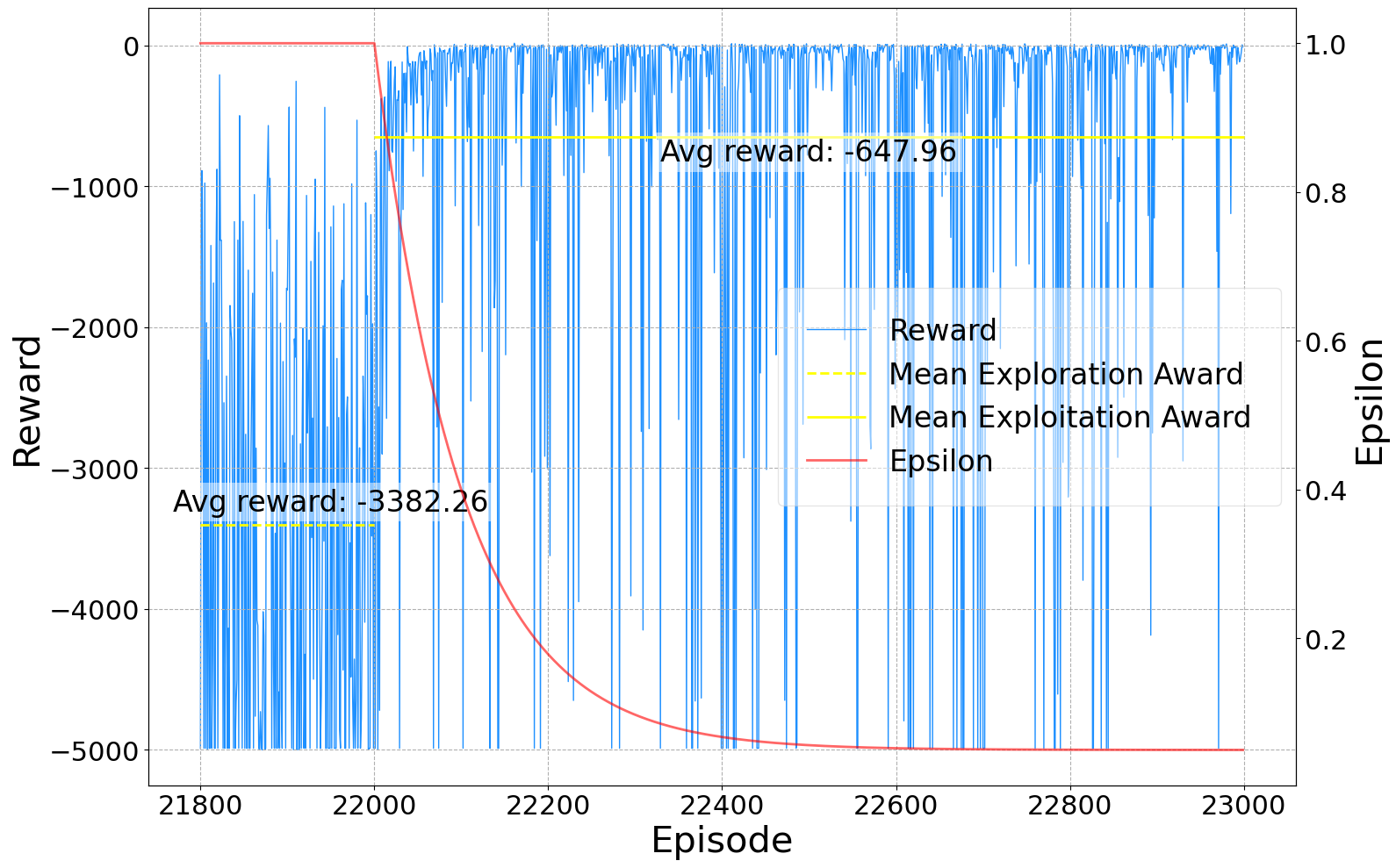}
        \label{fig:award-ddrqn-subset}
    }
    \hfill
    \subfigure[DRQN]{
        \includegraphics[width=.48\textwidth]{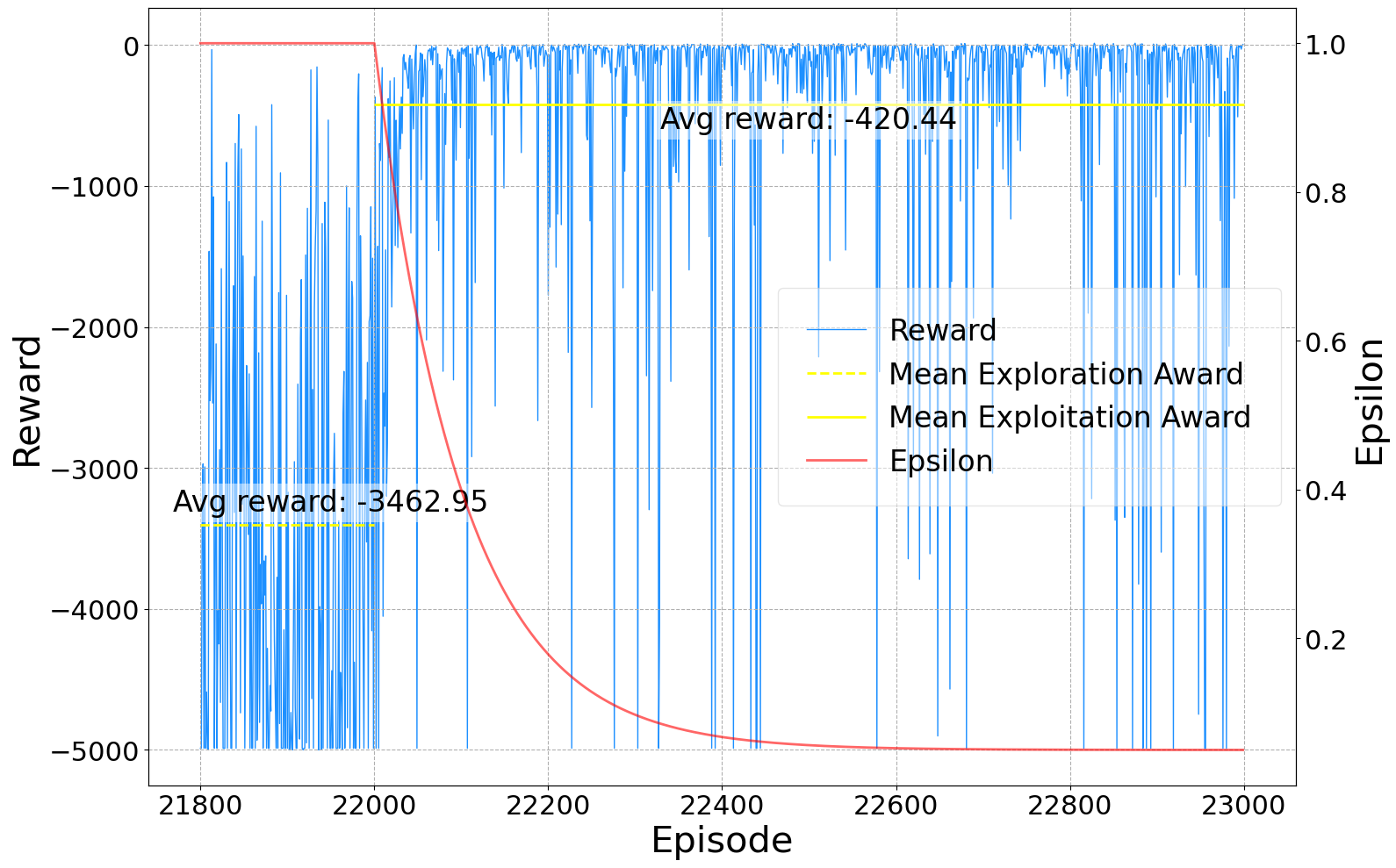}
        \label{fig:award-drqn-subset}
    }
  
    \caption{Learning Curves for learning-based models.}
    \label{fig:learning_curve}
\end{figure*}

The performance results outlined in Table \ref{tab:rl_model_performance} , coupled with the learning curves in Figure \ref{fig:learning_curve}, reveal several noteworthy findings.  First, all reinforcement learning (RL) models substantially outperform a random baseline, which, under the same experimental conditions, achieved a success rate below 8\%. Such a result underscores the importance of leveraging specialized learning algorithms over simpler stochastic approaches, particularly in scenarios where efficient exploration is paramount. Second, the incorporation of action-observation pairs in the learning architecture proves to be unequivocally superior in discovering latent states in partially observable environments. This is evidenced by the consistent outperformance of action-dependent models (ADDRQN and ADRQN) compared to their non-action-dependent counterparts (DDRQN and DRQN) in all KPIs. On average, action-dependent models achieve a 43.35\% higher success rate, require 35.18\% fewer epochs to reach a terminal state, and demonstrate a 41.90\% reduction in training time compared to their non-action-dependent counterparts. The learning curves in Figure 6 further illustrate this superiority, showing that ADDRQN and ADRQN exhibit more stable performance post-convergence, with fewer large fluctuations in rewards compared to DDRQN and DRQN. Finally, The combination of high success rate, low average epochs, high average reward, and fastest training time posits ADDRQN as the most balanced and efficient model among those tested. While both  ADDRQN and ADRQN models achieve high success rates, ADRQN demonstrates slightly higher consistency in successfully completing the localization tasks, with a success rate of 98.66\% compared to ADDRQN's 95.97\%. However, ADDRQN exhibits superior efficiency with  more optimal path-planning and decision-making. Namely, ADDRQN requires fewer epochs to reach a terminal state, accumulates higher average rewards per episode, and completes training in 21.39\% less time than ADRQN. Figure \ref{fig:addrqn_trajectory} illustrates selected sample of agent's trajectories maneuvering through the Gaussian Plume environments.

\subsection{Multi-Agent Experiments}
This section discusses and analyzes supplemental experiments, namely we investigate how varying the number of agents affects the performance of our ADDRQN algorithm. Specifically, we analyze changes in success rate, average epochs, training time, and overall reward efficiency.

\begin{table*}[!htb]
  \centering
  \caption{Performance summary of ADDQRN models with varying agent count}
  \begin{tabular}{lccccc}
    \toprule
    Metric & 1 & 2 & 3 & 4 & 5 \\
    \midrule
    Success Rate      & 90.08\%  & 95.97\%  & 99.34\% & 97.44\% & 98.66\%  \\
    Avg Epochs        & 74.78    & 52.06    & 44.63   & 50.19   &  47.54   \\
    Avg Reward        & -50.78   & -28.06   & -20.63  & -26.19  & -23.54    \\
    Training Time (d) & 3.24     &  3.16    & 2.96   & 3.55     &  3.65   \\
    \bottomrule
  \end{tabular} 
  \label{tab:rl_supplemental_performance}
\end{table*}

The results summarized in Table \ref{tab:rl_supplemental_performance} indicate a clear trend: as the number of agents increases, there is a notable improvement in several key performance metrics. Specifically, there is a substantial relative increase in success rate of nearly 9.5\% when moving from one agent to five agents. This trend is further supported by a 36.5\% reduction in the average number of epochs required to complete the task as the agent count increases. While there are slight variations in performance between configurations with 3, 4, and 5 agents, these differences are marginal and can largely be attributed to stochastic variability inherent in training. The most compelling aspect of these results is the stability in training time across different agent configurations. Despite the increase in agent count, the training time remains relatively consistent, varying only slightly from 3.24 days for one agent to 3.65 days for five agents. This stability suggests that introducing more agents does not significantly increase the computational burden. Instead, it appears that the additional agents lead to more efficient exploration and faster convergence, allowing the model to learn more effectively without a substantial increase in training overhead. 

\begin{figure}[!htb]
    \centering
    \subfigure[Agents = 1]{
        \includegraphics[width=.48\columnwidth]{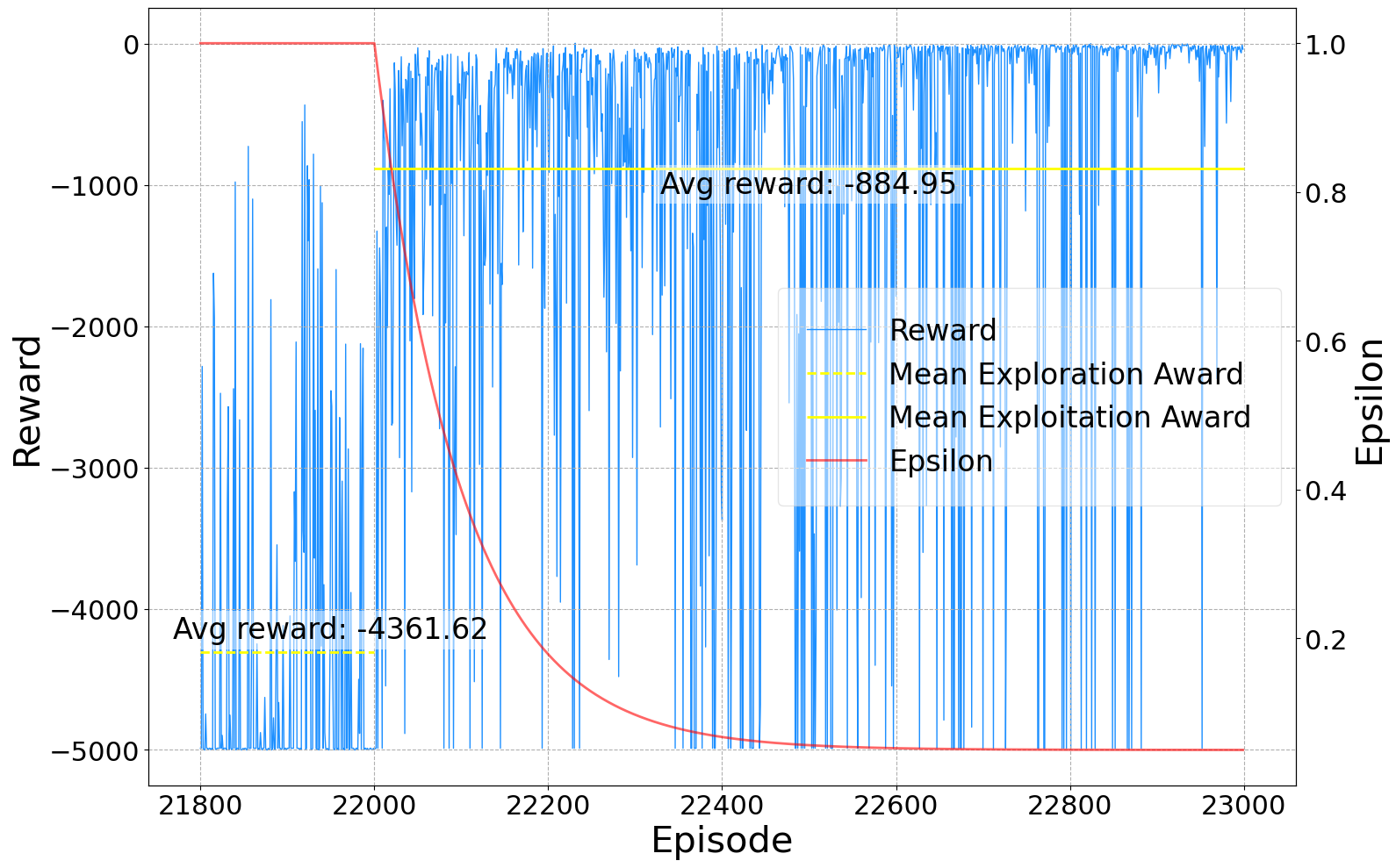}
        \label{fig:adrqn}
    }
    \hfil 
    \subfigure[Agent = 3]{
        \includegraphics[width=.48\columnwidth]{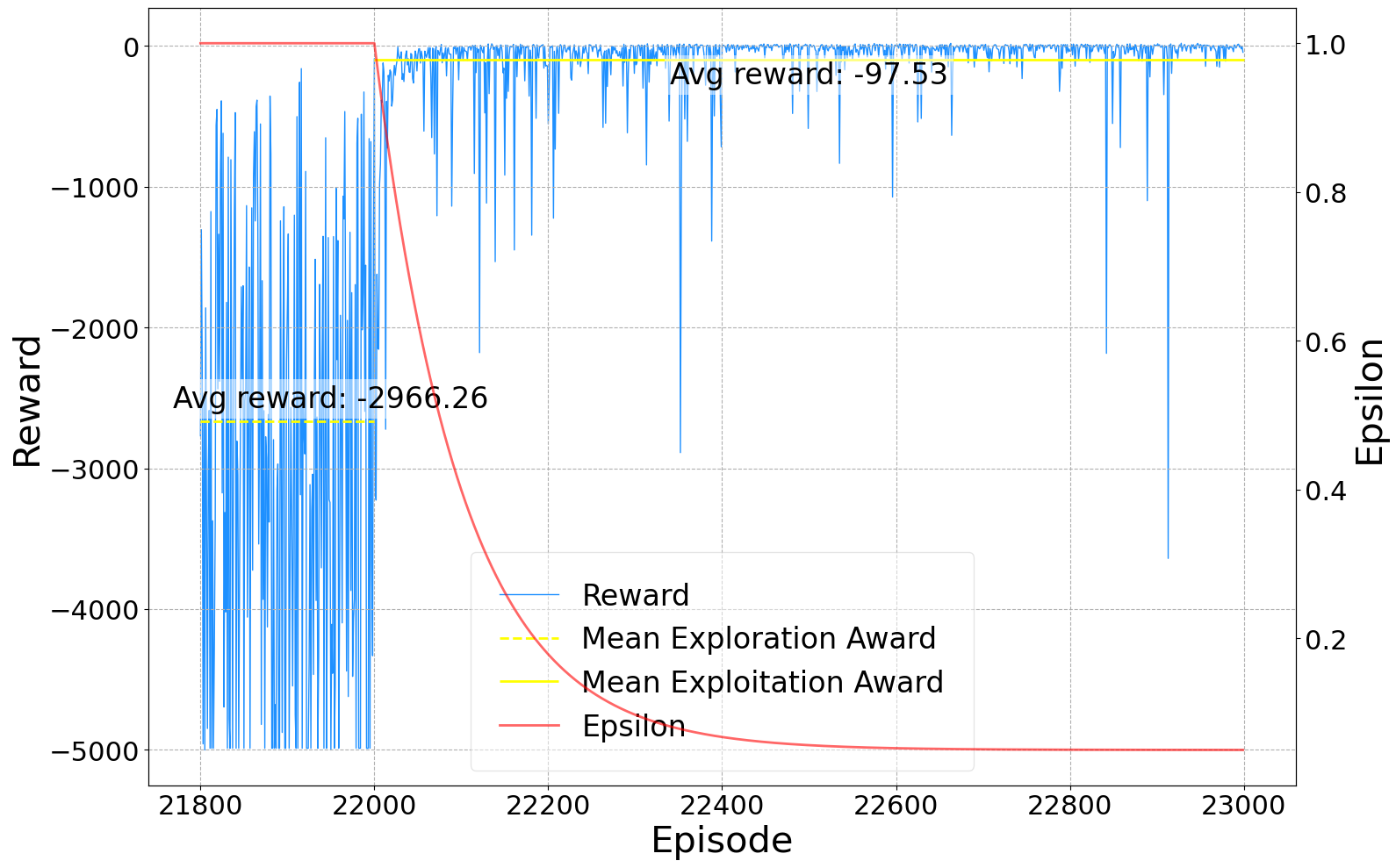}
        \label{fig:ddrqn}
    } \\
    \subfigure[Agent = 4]{
        \includegraphics[width=.48\columnwidth]{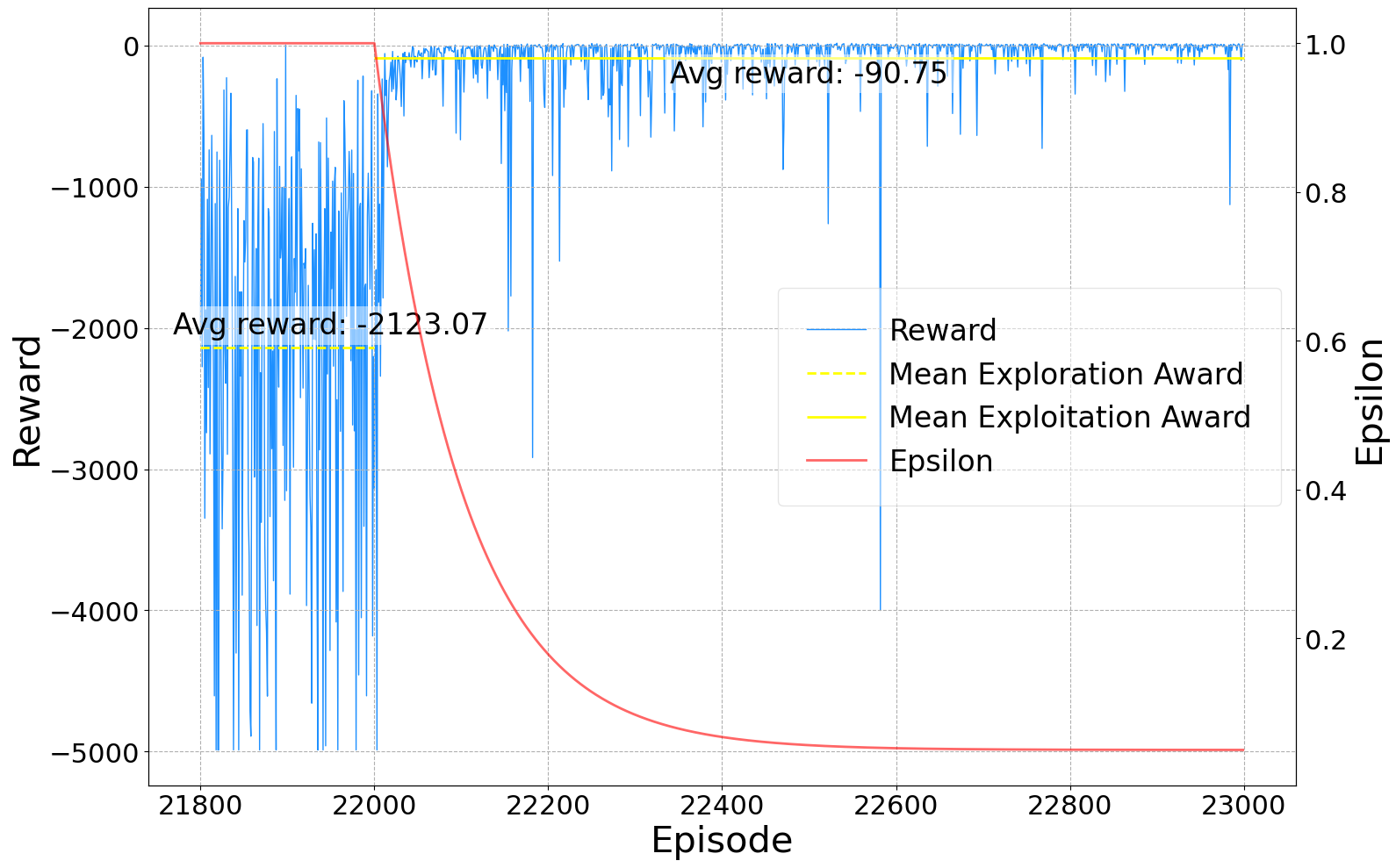}
        \label{fig:drqn}
    }
    \hfil 
    \subfigure[Agent = 5]{
        \includegraphics[width=.48\columnwidth]{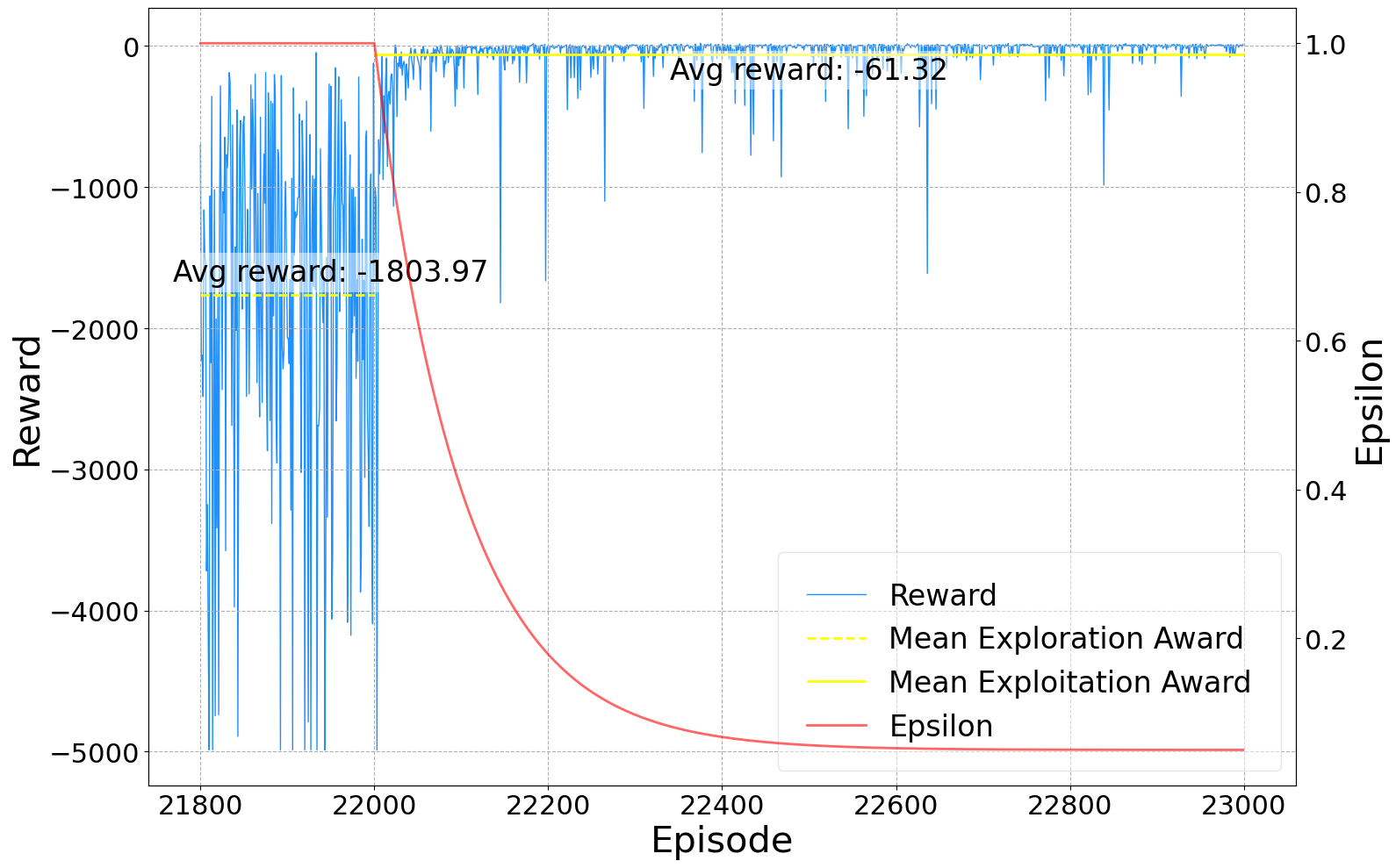}
        \label{fig:newalg}
    }
    \caption{Learning Curves of ADDRQN with varying number of agents}
    \label{fig:learning_curve_multiagent}
\end{figure}

The superior performance of multi-agent configurations can be attributed to two key factors. Figure \ref{fig:learning_curve_multiagent}'s learning curves show higher cumulative rewards during the exploration phase as the number of agents increases, indicating more effective area exploration. This improved coverage suggests that the presence of multiple agents enhances the likelihood of detecting pollution sources, even under stochastic exploration strategies. Second, the additional observational data from multiple agents enhances the model's ability to discern latent patterns in pollution distribution, leading to better coordination and decision-making. The learning curves in Figure \ref{fig:learning_curve_multiagent} further illustrate these trends. For single-agent setups, there is considerable instability in the early episodes, with high reward fluctuations and a slower convergence rate. In contrast, multi-agent setups, particularly with 3 or more agents, demonstrate more stable learning curves with fewer spikes and smoother convergence.

\section{Limitations and Future Directions} \label{sec:limitations}

The advantages of the Gaussian Dispersion Model (GDM) include its simplicity, computational efficiency, and effectiveness in short to medium-range dispersion (up to 20 kilometers) over flat, homogeneous terrains with continuous emissions under neutral and stable atmospheric conditions. It is widely recognized for simulating common air pollutants such as sulfur dioxide (SO\textsubscript{2}), nitrogen oxides (NO\textsubscript{x}), particulate matter (PM), carbon monoxide (CO), volatile organic compounds (VOCs), ammonia (NH\textsubscript{3}), and hydrogen sulfide (H\textsubscript{2}S). However, GDM has notable limitations, including reduced accuracy in complex terrains and urban environments, limited applicability to long-range transport and secondary pollutants like ozone (O\textsubscript{3}) and secondary VOCs \citep{Snoun2023}, and potential inaccuracies when applied to specific gases like CO2 without optimized dispersion parameters \citep{Liu2015}. These limitations necessitate careful consideration when choosing GDM for specific environmental modeling applications.

While the proposed Multi-Agent Reinforcement Learning (MARL) model efficiently addresses the plume detection and traversal phases, it does not encompass the source identification phase. One promising avenue is the incorporation of computer vision object detection models that can identify and confirm the characteristics of the source based on visual imagery \citep{wang2020machine}. Furthermore, the current MARL framework does not address operational challenges specific to UAVs, such as battery life, hardware failure, communication range, and obstacle avoidance. Future enhancements could include penalties for near-collisions to ensure safe UAV separation \citep{DUGULEANA2016104}, and using dropout methods or data augmentation to improve resilience to missing data and maintain effectiveness during UAV failures or communication disruptions. \citep{macielpearson2019onlinedeepreinforcementlearning}. Additionally, Robust Contrastive Learning (RoCL) \citep{10242114} offers a promising method to address noisy values and improve generalizability. RoCL could be integrated as a potential surrogate for the Gaussian noise protocol or as a complementary technique to further enhance the robustness and performance of our models.

\section{Conclusion}
\label{sec:conclusions}

In this work, we explored the  multi-source plume tracing problem in three-dimensional space. We present a model-free reinforcement learning solution based on the ADDRQN algorithm, contextualized as a Partially Observable Markov Game in a cooperative multi-agent setting. Our methodology includes a careful design of action and observation spaces, along with a targeted reward function. We evaluated our approach against well-established models such as ADRQN, DRQN, and DDRQN. Through the incorporation of Long Short-Term Memory (LSTM) units, our model provides a seamless integration of action-observation history, bridging the gap between $Q(s, a)$ and $Q(o, a)$. We used the Gaussian Plume Dispersion Model to create a realistic 3D environment for testing the benefits of multiple agents in multi-source localization. Our primary experiments have shown ADDRQN's exceptional ability in identifying pollution sources with high accuracy and efficiency, outperforming its counterparts in success rate, average epochs, and reward accumulation. Furthermore, our supplemental experiments demonstrate that increasing the number of agents in the ADDRQN framework significantly enhances performance across various metrics, including success rate, average epochs, and reward efficiency. This improvement can be attributed to a combination of wider area coverage, enhanced data input for improved decision-making, and more consistent learning processes in multi-agent configurations. These findings not only validate the scalability of our approach but also highlight the potential advantages of employing multi-agent systems in complex tasks like pollution source localization.

\section*{Acknowledgment}
The work in this paper has been funded by the US National Science Foundation (NSF) under Grant\# CNS-1931962.





\begin{thebibliography}{37}
\expandafter\ifx\csname natexlab\endcsname\relax\def\natexlab#1{#1}\fi
\providecommand{\url}[1]{\texttt{#1}}
\providecommand{\href}[2]{#2}
\providecommand{\path}[1]{#1}
\providecommand{\DOIprefix}{doi:}
\providecommand{\ArXivprefix}{arXiv:}
\providecommand{\URLprefix}{URL: }
\providecommand{\Pubmedprefix}{pmid:}
\providecommand{\doi}[1]{\href{http://dx.doi.org/#1}{\path{#1}}}
\providecommand{\Pubmed}[1]{\href{pmid:#1}{\path{#1}}}
\providecommand{\bibinfo}[2]{#2}
\ifx\xfnm\relax \def\xfnm[#1]{\unskip,\space#1}\fi
\bibitem[{Alvear et~al.(2018)Alvear, Calafate, Zema, Natalizio, Hern{\'a}ndez-Orallo, Cano and Manzoni}]{Alvear2018}
\bibinfo{author}{Alvear, O.}, \bibinfo{author}{Calafate, C.T.}, \bibinfo{author}{Zema, N.R.}, \bibinfo{author}{Natalizio, E.}, \bibinfo{author}{Hern{\'a}ndez-Orallo, E.}, \bibinfo{author}{Cano, J.C.}, \bibinfo{author}{Manzoni, P.}, \bibinfo{year}{2018}.
\newblock \bibinfo{title}{A discretized approach to air pollution monitoring using uav-based sensing}.
\newblock \bibinfo{journal}{Mobile Networks and Applications} \bibinfo{volume}{23}, \bibinfo{pages}{1693--1702}.
\newblock \URLprefix \url{https://doi.org/10.1007/s11036-018-1065-4}, \DOIprefix\doi{10.1007/s11036-018-1065-4}.
\bibitem[{Bayat et~al.(2017)Bayat, Crasta, Crespi, Pascoal and Ijspeert}]{BAYAT201776}
\bibinfo{author}{Bayat, B.}, \bibinfo{author}{Crasta, N.}, \bibinfo{author}{Crespi, A.}, \bibinfo{author}{Pascoal, A.M.}, \bibinfo{author}{Ijspeert, A.}, \bibinfo{year}{2017}.
\newblock \bibinfo{title}{Environmental monitoring using autonomous vehicles: a survey of recent searching techniques}.
\newblock \bibinfo{journal}{Current Opinion in Biotechnology} \bibinfo{volume}{45}, \bibinfo{pages}{76--84}.
\newblock \URLprefix \url{https://www.sciencedirect.com/science/article/pii/S0958166916302312}, \DOIprefix\doi{https://doi.org/10.1016/j.copbio.2017.01.009}. \bibinfo{note}{energy biotechnology • Environmental biotechnology}.
\bibitem[{Broughton(2005)}]{Broughton2005}
\bibinfo{author}{Broughton, E.}, \bibinfo{year}{2005}.
\newblock \bibinfo{title}{The bhopal disaster and its aftermath: a review}.
\newblock \bibinfo{journal}{Environmental Health} \bibinfo{volume}{4}, \bibinfo{pages}{6}.
\newblock \URLprefix \url{https://doi.org/10.1186/1476-069X-4-6}, \DOIprefix\doi{10.1186/1476-069X-4-6}.
\bibitem[{Casbeer et~al.(2005)Casbeer, Beard, McLain, Li and Mehra}]{1470520}
\bibinfo{author}{Casbeer, D.}, \bibinfo{author}{Beard, R.}, \bibinfo{author}{McLain, T.}, \bibinfo{author}{Li, S.M.}, \bibinfo{author}{Mehra, R.}, \bibinfo{year}{2005}.
\newblock \bibinfo{title}{Forest fire monitoring with multiple small uavs}, in: \bibinfo{booktitle}{Proceedings of the 2005, American Control Conference, 2005.}, pp. \bibinfo{pages}{3530--3535 vol. 5}.
\newblock \DOIprefix\doi{10.1109/ACC.2005.1470520}.
\bibitem[{Chastain and Wolak(2000)}]{LIVESTOCKPLUME}
\bibinfo{author}{Chastain, J.}, \bibinfo{author}{Wolak, F.}, \bibinfo{year}{2000}.
\newblock \bibinfo{title}{Application of a gaussian plume model of odor dispersion to select a site for livestock facilities}.
\newblock \bibinfo{journal}{Proceedings of the Water Environment Federation} \bibinfo{volume}{2000}, \bibinfo{pages}{745--758}.
\newblock \DOIprefix\doi{10.2175/193864700785303321}.
\bibitem[{Clark and Fierro(2005)}]{1470515}
\bibinfo{author}{Clark, J.}, \bibinfo{author}{Fierro, R.}, \bibinfo{year}{2005}.
\newblock \bibinfo{title}{Cooperative hybrid control of robotic sensors for perimeter detection and tracking}, in: \bibinfo{booktitle}{Proceedings of the 2005, American Control Conference, 2005.}, pp. \bibinfo{pages}{3500--3505 vol. 5}.
\newblock \DOIprefix\doi{10.1109/ACC.2005.1470515}.
\bibitem[{Conley et~al.(2016)Conley, Franco, Faloona, Blake, Peischl and Ryerson}]{doi:10.1126/science.aaf2348}
\bibinfo{author}{Conley, S.}, \bibinfo{author}{Franco, G.}, \bibinfo{author}{Faloona, I.}, \bibinfo{author}{Blake, D.R.}, \bibinfo{author}{Peischl, J.}, \bibinfo{author}{Ryerson, T.B.}, \bibinfo{year}{2016}.
\newblock \bibinfo{title}{Methane emissions from the 2015 aliso canyon blowout in los angeles, ca}.
\newblock \bibinfo{journal}{Science} \bibinfo{volume}{351}, \bibinfo{pages}{1317--1320}.
\newblock \URLprefix \url{https://www.science.org/doi/abs/10.1126/science.aaf2348}, \DOIprefix\doi{10.1126/science.aaf2348}, \href{http://arxiv.org/abs/https://www.science.org/doi/pdf/10.1126/science.aaf2348}{{\tt arXiv:https://www.science.org/doi/pdf/10.1126/science.aaf2348}}.
\bibitem[{Duguleana and Mogan(2016)}]{DUGULEANA2016104}
\bibinfo{author}{Duguleana, M.}, \bibinfo{author}{Mogan, G.}, \bibinfo{year}{2016}.
\newblock \bibinfo{title}{Neural networks based reinforcement learning for mobile robots obstacle avoidance}.
\newblock \bibinfo{journal}{Expert Systems with Applications} \bibinfo{volume}{62}, \bibinfo{pages}{104--115}.
\newblock \URLprefix \url{https://www.sciencedirect.com/science/article/pii/S0957417416303001}, \DOIprefix\doi{https://doi.org/10.1016/j.eswa.2016.06.021}.
\bibitem[{Farrell et~al.(2005)Farrell, Pang and Li}]{1522521}
\bibinfo{author}{Farrell, J.}, \bibinfo{author}{Pang, S.}, \bibinfo{author}{Li, W.}, \bibinfo{year}{2005}.
\newblock \bibinfo{title}{Chemical plume tracing via an autonomous underwater vehicle}.
\newblock \bibinfo{journal}{IEEE Journal of Oceanic Engineering} \bibinfo{volume}{30}, \bibinfo{pages}{428--442}.
\newblock \DOIprefix\doi{10.1109/JOE.2004.838066}.
\bibitem[{Graves(2012)}]{Graves2012}
\bibinfo{author}{Graves, A.}, \bibinfo{year}{2012}.
\newblock \bibinfo{title}{Long Short-Term Memory}. \bibinfo{publisher}{Springer Berlin Heidelberg}, \bibinfo{address}{Berlin, Heidelberg}.
\newblock pp. \bibinfo{pages}{37--45}.
\newblock \URLprefix \url{https://doi.org/10.1007/978-3-642-24797-2_4}, \DOIprefix\doi{10.1007/978-3-642-24797-2_4}.
\bibitem[{Harper and Dock(2007)}]{10.1117/12.719742}
\bibinfo{author}{Harper, R.J.}, \bibinfo{author}{Dock, M.L.}, \bibinfo{year}{2007}.
\newblock \bibinfo{title}{{Underwater olfaction for real-time detection of submerged unexploded ordnance}}, in: \bibinfo{editor}{Saito, T.T.}, \bibinfo{editor}{Lehrfeld, D.}, \bibinfo{editor}{DeWeert, M.J.} (Eds.), \bibinfo{booktitle}{Optics and Photonics in Global Homeland Security III}, \bibinfo{organization}{International Society for Optics and Photonics}. \bibinfo{publisher}{SPIE}. p. \bibinfo{pages}{65400V}.
\newblock \URLprefix \url{https://doi.org/10.1117/12.719742}, \DOIprefix\doi{10.1117/12.719742}.
\bibitem[{van Hasselt et~al.(2015)van Hasselt, Guez and Silver}]{vanhasselt2015deep}
\bibinfo{author}{van Hasselt, H.}, \bibinfo{author}{Guez, A.}, \bibinfo{author}{Silver, D.}, \bibinfo{year}{2015}.
\newblock \bibinfo{title}{Deep reinforcement learning with double q-learning}.
\newblock \href{http://arxiv.org/abs/1509.06461}{{\tt arXiv:1509.06461}}.
\bibitem[{Hayes et~al.(2002)Hayes, Martinoli and Goodman}]{1021067}
\bibinfo{author}{Hayes, A.}, \bibinfo{author}{Martinoli, A.}, \bibinfo{author}{Goodman, R.}, \bibinfo{year}{2002}.
\newblock \bibinfo{title}{Distributed odor source localization}.
\newblock \bibinfo{journal}{IEEE Sensors Journal} \bibinfo{volume}{2}, \bibinfo{pages}{260--271}.
\newblock \DOIprefix\doi{10.1109/JSEN.2002.800682}.
\bibitem[{Hu et~al.(2019)Hu, Song and Chen}]{8598800}
\bibinfo{author}{Hu, H.}, \bibinfo{author}{Song, S.}, \bibinfo{author}{Chen, C.L.P.}, \bibinfo{year}{2019}.
\newblock \bibinfo{title}{Plume tracing via model-free reinforcement learning method}.
\newblock \bibinfo{journal}{IEEE Transactions on Neural Networks and Learning Systems} \bibinfo{volume}{30}, \bibinfo{pages}{2515--2527}.
\newblock \DOIprefix\doi{10.1109/TNNLS.2018.2885374}.
\bibitem[{Kurniawati(2022)}]{annurev:/content/journals/10.1146/annurev-control-042920-092451}
\bibinfo{author}{Kurniawati, H.}, \bibinfo{year}{2022}.
\newblock \bibinfo{title}{Partially observable markov decision processes and robotics}.
\newblock \bibinfo{journal}{Annual Review of Control, Robotics, and Autonomous Systems} \bibinfo{volume}{5}, \bibinfo{pages}{253--277}.
\newblock \URLprefix \url{https://www.annualreviews.org/content/journals/10.1146/annurev-control-042920-092451}, \DOIprefix\doi{https://doi.org/10.1146/annurev-control-042920-092451}.
\bibitem[{Li et~al.(2006)Li, Farrell, Pang and Arrieta}]{1618529}
\bibinfo{author}{Li, W.}, \bibinfo{author}{Farrell, J.}, \bibinfo{author}{Pang, S.}, \bibinfo{author}{Arrieta, R.}, \bibinfo{year}{2006}.
\newblock \bibinfo{title}{Moth-inspired chemical plume tracing on an autonomous underwater vehicle}.
\newblock \bibinfo{journal}{IEEE Transactions on Robotics} \bibinfo{volume}{22}, \bibinfo{pages}{292--307}.
\newblock \DOIprefix\doi{10.1109/TRO.2006.870627}.
\bibitem[{Liu et~al.(2015)Liu, Godbole, Lu, Michal and Venton}]{Liu2015}
\bibinfo{author}{Liu, X.}, \bibinfo{author}{Godbole, A.}, \bibinfo{author}{Lu, C.}, \bibinfo{author}{Michal, G.}, \bibinfo{author}{Venton, P.}, \bibinfo{year}{2015}.
\newblock \bibinfo{title}{Optimisation of dispersion parameters of {Gaussian} plume model for {CO} dispersion}.
\newblock \bibinfo{journal}{Environmental Science and Pollution Research} \bibinfo{volume}{22}, \bibinfo{pages}{18288--18299}.
\newblock \DOIprefix\doi{10.1007/s11356-015-5404-8}. \bibinfo{note}{pMID: 26374541}.
\bibitem[{Maciel-Pearson et~al.(2019)Maciel-Pearson, Marchegiani, Akcay, Atapour-Abarghouei, Garforth and Breckon}]{macielpearson2019onlinedeepreinforcementlearning}
\bibinfo{author}{Maciel-Pearson, B.G.}, \bibinfo{author}{Marchegiani, L.}, \bibinfo{author}{Akcay, S.}, \bibinfo{author}{Atapour-Abarghouei, A.}, \bibinfo{author}{Garforth, J.}, \bibinfo{author}{Breckon, T.P.}, \bibinfo{year}{2019}.
\newblock \bibinfo{title}{Online deep reinforcement learning for autonomous uav navigation and exploration of outdoor environments}.
\newblock \URLprefix \url{https://arxiv.org/abs/1912.05684}, \href{http://arxiv.org/abs/1912.05684}{{\tt arXiv:1912.05684}}.
\bibitem[{Mayhew et~al.(2007)Mayhew, Sanfelice and Teel}]{12}
\bibinfo{author}{Mayhew, C.G.}, \bibinfo{author}{Sanfelice, R.G.}, \bibinfo{author}{Teel, A.R.}, \bibinfo{year}{2007}.
\newblock \bibinfo{title}{Robust source seeking hybrid controllers for autonomous vehicles}, in: \bibinfo{booktitle}{Proc. 26th American Control Conference}, p. \bibinfo{pages}{1185{\textendash}1190}.
\newblock \URLprefix \url{https://hybrid.soe.ucsc.edu/files/preprints/12.pdf}, \DOIprefix\doi{http://www.scivee.tv/node/2725,http\://ieeexplore.ieee.org/iel5/4282134/4282135/04283016.pdf?tp=\&isnumber=4282135\&arnumber=4283016\&punumber=\%3Cb\%3E\%3Cfont\%20color=990000\%3E4282134\%3C/font\%3E\%3C/b\%3E}.
\bibitem[{Middleton et~al.(1979)Middleton, Butler and Colwill}]{MIDDLETON19791039}
\bibinfo{author}{Middleton, D.}, \bibinfo{author}{Butler, J.}, \bibinfo{author}{Colwill, D.}, \bibinfo{year}{1979}.
\newblock \bibinfo{title}{Gaussian plume dispersion model applicable to a complex motorway interchange}.
\newblock \bibinfo{journal}{Atmospheric Environment (1967)} \bibinfo{volume}{13}, \bibinfo{pages}{1039--1049}.
\newblock \URLprefix \url{https://www.sciencedirect.com/science/article/pii/0004698179900143}, \DOIprefix\doi{https://doi.org/10.1016/0004-6981(79)90014-3}.
\bibitem[{Mnih et~al.(2013)Mnih, Kavukcuoglu, Silver, Graves, Antonoglou, Wierstra and Riedmiller}]{mnih2013playing}
\bibinfo{author}{Mnih, V.}, \bibinfo{author}{Kavukcuoglu, K.}, \bibinfo{author}{Silver, D.}, \bibinfo{author}{Graves, A.}, \bibinfo{author}{Antonoglou, I.}, \bibinfo{author}{Wierstra, D.}, \bibinfo{author}{Riedmiller, M.}, \bibinfo{year}{2013}.
\newblock \bibinfo{title}{Playing atari with deep reinforcement learning}.
\newblock \href{http://arxiv.org/abs/1312.5602}{{\tt arXiv:1312.5602}}.
\bibitem[{Mohammed et~al.(2022)Mohammed, Sultan, Cho and Pyun}]{s22166118}
\bibinfo{author}{Mohammed, A.F.Y.}, \bibinfo{author}{Sultan, S.M.}, \bibinfo{author}{Cho, S.}, \bibinfo{author}{Pyun, J.Y.}, \bibinfo{year}{2022}.
\newblock \bibinfo{title}{Powering uav with deep q-network for air quality tracking}.
\newblock \bibinfo{journal}{Sensors} \bibinfo{volume}{22}.
\newblock \URLprefix \url{https://www.mdpi.com/1424-8220/22/16/6118}, \DOIprefix\doi{10.3390/s22166118}.
\bibitem[{Neumann et~al.(2013)Neumann, Bennetts, Lilienthal, Bartholmai and Schiller}]{GasSourceMicro}
\bibinfo{author}{Neumann, P.P.}, \bibinfo{author}{Bennetts, V.H.}, \bibinfo{author}{Lilienthal, A.J.}, \bibinfo{author}{Bartholmai, M.}, \bibinfo{author}{Schiller, J.H.}, \bibinfo{year}{2013}.
\newblock \bibinfo{title}{Gas source localization with a micro-drone using bio-inspired and particle filter-based algorithms}.
\newblock \bibinfo{journal}{Advanced Robotics} \bibinfo{volume}{27}, \bibinfo{pages}{725--738}.
\newblock \URLprefix \url{https://doi.org/10.1080/01691864.2013.779052}, \DOIprefix\doi{10.1080/01691864.2013.779052}, \href{http://arxiv.org/abs/https://doi.org/10.1080/01691864.2013.779052}{{\tt arXiv:https://doi.org/10.1080/01691864.2013.779052}}.
\bibitem[{Puterman(1990)}]{PUTERMAN1990331}
\bibinfo{author}{Puterman, M.L.}, \bibinfo{year}{1990}.
\newblock \bibinfo{title}{Chapter 8 markov decision processes}, in: \bibinfo{booktitle}{Stochastic Models}. \bibinfo{publisher}{Elsevier}. volume~\bibinfo{volume}{2} of \textit{\bibinfo{series}{Handbooks in Operations Research and Management Science}}, pp. \bibinfo{pages}{331--434}.
\newblock \URLprefix \url{https://www.sciencedirect.com/science/article/pii/S0927050705801720}, \DOIprefix\doi{https://doi.org/10.1016/S0927-0507(05)80172-0}.
\bibitem[{Silver et~al.(2016)Silver, Huang, Maddison et~al.}]{Silver2016}
\bibinfo{author}{Silver, D.}, \bibinfo{author}{Huang, A.}, \bibinfo{author}{Maddison, C.}, et~al., \bibinfo{year}{2016}.
\newblock \bibinfo{title}{Mastering the game of go with deep neural networks and tree search}.
\newblock \bibinfo{journal}{Nature} \bibinfo{volume}{529}, \bibinfo{pages}{484--489}.
\newblock \DOIprefix\doi{10.1038/nature16961}.
\bibitem[{Silver et~al.(2017)Silver, Schrittwieser, Simonyan et~al.}]{Silver2017}
\bibinfo{author}{Silver, D.}, \bibinfo{author}{Schrittwieser, J.}, \bibinfo{author}{Simonyan, K.}, et~al., \bibinfo{year}{2017}.
\newblock \bibinfo{title}{Mastering the game of go without human knowledge}.
\newblock \bibinfo{journal}{Nature} \bibinfo{volume}{550}, \bibinfo{pages}{354--359}.
\newblock \DOIprefix\doi{10.1038/nature24270}.
\bibitem[{Snoun et~al.(2023)Snoun, Krichen and Chérif}]{Snoun2023}
\bibinfo{author}{Snoun, H.}, \bibinfo{author}{Krichen, M.}, \bibinfo{author}{Chérif, H.}, \bibinfo{year}{2023}.
\newblock \bibinfo{title}{A comprehensive review of gaussian atmospheric dispersion models: current usage and future perspectives}.
\newblock \bibinfo{journal}{Euro-Mediterranean Journal for Environmental Integration} \bibinfo{volume}{8}, \bibinfo{pages}{219--242}.
\newblock \URLprefix \url{https://doi.org/10.1007/s41207-023-00354-6}, \DOIprefix\doi{10.1007/s41207-023-00354-6}.
\bibitem[{Stockie(2011)}]{doi:10.1137/10080991X}
\bibinfo{author}{Stockie, J.M.}, \bibinfo{year}{2011}.
\newblock \bibinfo{title}{The mathematics of atmospheric dispersion modeling}.
\newblock \bibinfo{journal}{SIAM Review} \bibinfo{volume}{53}, \bibinfo{pages}{349--372}.
\newblock \URLprefix \url{https://doi.org/10.1137/10080991X}, \DOIprefix\doi{10.1137/10080991X}, \href{http://arxiv.org/abs/https://doi.org/10.1137/10080991X}{{\tt arXiv:https://doi.org/10.1137/10080991X}}.
\bibitem[{Tate et~al.(2021)Tate, Fries, Vignati and Francis}]{9706111}
\bibinfo{author}{Tate, C.}, \bibinfo{author}{Fries, D.P.}, \bibinfo{author}{Vignati, M.}, \bibinfo{author}{Francis, K.}, \bibinfo{year}{2021}.
\newblock \bibinfo{title}{Using model-free reinforcement learning combined with underwater mass spectrometer and material archiving coupled to lab analysis for autonomous chemical source verifications}, in: \bibinfo{booktitle}{OCEANS 2021: San Diego – Porto}, pp. \bibinfo{pages}{1--9}.
\newblock \DOIprefix\doi{10.23919/OCEANS44145.2021.9706111}.
\bibitem[{Tsitsiklis and Van~Roy(1997)}]{580874}
\bibinfo{author}{Tsitsiklis, J.}, \bibinfo{author}{Van~Roy, B.}, \bibinfo{year}{1997}.
\newblock \bibinfo{title}{An analysis of temporal-difference learning with function approximation}.
\newblock \bibinfo{journal}{IEEE Transactions on Automatic Control} \bibinfo{volume}{42}, \bibinfo{pages}{674--690}.
\newblock \DOIprefix\doi{10.1109/9.580874}.
\bibitem[{Veigele and Head(1978)}]{DerivGaussianVeigele}
\bibinfo{author}{Veigele, W.J.}, \bibinfo{author}{Head, J.H.}, \bibinfo{year}{1978}.
\newblock \bibinfo{title}{Derivation of the gaussian plume model}.
\newblock \bibinfo{journal}{Journal of the Air Pollution Control Association} \bibinfo{volume}{28}, \bibinfo{pages}{1139--1140}.
\newblock \URLprefix \url{https://doi.org/10.1080/00022470.1978.10470720}, \DOIprefix\doi{10.1080/00022470.1978.10470720}, \href{http://arxiv.org/abs/https://doi.org/10.1080/00022470.1978.10470720}{{\tt arXiv:https://doi.org/10.1080/00022470.1978.10470720}}.
\bibitem[{Wang and Hu(2023)}]{10242114}
\bibinfo{author}{Wang, D.}, \bibinfo{author}{Hu, M.}, \bibinfo{year}{2023}.
\newblock \bibinfo{title}{Contrastive learning methods for deep reinforcement learning}.
\newblock \bibinfo{journal}{IEEE Access} \bibinfo{volume}{11}, \bibinfo{pages}{97107--97117}.
\newblock \DOIprefix\doi{10.1109/ACCESS.2023.3312383}.
\bibitem[{Wang et~al.(2020)Wang, Tchapmi, Ravikumar, McGuire, Bell, Zimmerle, Savarese and Brandt}]{wang2020machine}
\bibinfo{author}{Wang, J.}, \bibinfo{author}{Tchapmi, L.P.}, \bibinfo{author}{Ravikumar, A.P.}, \bibinfo{author}{McGuire, M.}, \bibinfo{author}{Bell, C.S.}, \bibinfo{author}{Zimmerle, D.}, \bibinfo{author}{Savarese, S.}, \bibinfo{author}{Brandt, A.R.}, \bibinfo{year}{2020}.
\newblock \bibinfo{title}{Machine vision for natural gas methane emissions detection using an infrared camera}.
\newblock \bibinfo{journal}{Applied Energy} \bibinfo{volume}{257}, \bibinfo{pages}{113998}.
\bibitem[{Wang and Wu(2020)}]{WangGasMulti2d}
\bibinfo{author}{Wang, Z.P.}, \bibinfo{author}{Wu, H.N.}, \bibinfo{year}{2020}.
\newblock \bibinfo{title}{Gas source localization using improved multi-agent reinforcement learning}, in: \bibinfo{booktitle}{2020 Chinese Automation Congress (CAC)}, pp. \bibinfo{pages}{6696--6701}.
\newblock \DOIprefix\doi{10.1109/CAC51589.2020.9327850}.
\bibitem[{Yungaicela-Naula et~al.(2018)Yungaicela-Naula, Zhang, Garza-Castañon and Minchala}]{8453430}
\bibinfo{author}{Yungaicela-Naula, N.M.}, \bibinfo{author}{Zhang, Y.}, \bibinfo{author}{Garza-Castañon, L.E.}, \bibinfo{author}{Minchala, L.I.}, \bibinfo{year}{2018}.
\newblock \bibinfo{title}{Uav-based air pollutant source localization using gradient and probabilistic methods}, in: \bibinfo{booktitle}{2018 International Conference on Unmanned Aircraft Systems (ICUAS)}, pp. \bibinfo{pages}{702--707}.
\newblock \DOIprefix\doi{10.1109/ICUAS.2018.8453430}.
\bibitem[{Zhang et~al.(2021)Zhang, Yang and Başar}]{zhang2021multiagentreinforcementlearningselective}
\bibinfo{author}{Zhang, K.}, \bibinfo{author}{Yang, Z.}, \bibinfo{author}{Başar, T.}, \bibinfo{year}{2021}.
\newblock \bibinfo{title}{Multi-agent reinforcement learning: A selective overview of theories and algorithms}.
\newblock \URLprefix \url{https://arxiv.org/abs/1911.10635}, \href{http://arxiv.org/abs/1911.10635}{{\tt arXiv:1911.10635}}.
\bibitem[{Zhu et~al.(2018)Zhu, Li, Poupart and Miao}]{zhu2018improving}
\bibinfo{author}{Zhu, P.}, \bibinfo{author}{Li, X.}, \bibinfo{author}{Poupart, P.}, \bibinfo{author}{Miao, G.}, \bibinfo{year}{2018}.
\newblock \bibinfo{title}{On improving deep reinforcement learning for pomdps}.
\newblock \href{http://arxiv.org/abs/1704.07978}{{\tt arXiv:1704.07978}}.

\end{thebibliography}
\end{document}